\documentclass[aps,prl,twocolumn,amssymb,superscriptaddress]{revtex4-2}
\usepackage{subfigure}
\usepackage{graphicx}
\usepackage{sidecap}
\usepackage{rotating} 
\usepackage{color}
\usepackage{rotating}
 \usepackage{amsmath,amsthm}
 \usepackage{enumitem}
\usepackage{epstopdf}
\begin{document}
\title{Irregularity of polymer domain boundaries in two-dimensional polymer solution}
\author{Lei Liu}
\affiliation{Key Laboratory of Optical Field Manipulation of Zhejiang Province, Department of Physics, Zhejiang Sci-Tech University, Hangzhou 310018, China}
\author{Changbong Hyeon}
\thanks{hyeoncb@kias.re.kr}
\affiliation{Korea Institute for Advanced Study, Seoul 02455, Korea}
\date{\today}

\begin{abstract}
Polymer chains {\color{black}composing} a polymer solution in strict two dimensions (2D) are characterized with irregular domain boundaries, whose fractal dimension ($\mathcal{D}^{\partial}$) varies with the area fraction of the solution and the solvent quality. 
{\color{black}Our analysis of numerical simulations of polymer solutions finds} that $\mathcal{D}^{\partial}$ in good solvents changes non-monotonically from $\mathcal{D}^{\partial}=4/3$ in dilute phase to $\mathcal{D}^{\partial}=5/4$ in dense phase, maximizing to $\mathcal{D}^{\partial}\approx 3/2$ at a {\color{black}crossover area fraction $\phi_{\rm cr}\approx 0.2$}, whereas for polymers in $\Theta$ solvents $\mathcal{D}^{\partial}$ remains constant at $\mathcal{D}^{\partial}=4/3$ from dilute to semi-dilute phase. 
Using polymer physics arguments, we rationalize these values, and show that the maximum irregularity of $\mathcal{D}^\partial\approx 3/2$ is due to ``fjord''-like corrugations formed along the domain boundaries which also maximize at the same {\color{black}crossover} area fraction. 
Our finding of $\mathcal{D}^\partial\approx 3/2$ is, in fact, in perfect agreement with the upper bound for the fractal dimension of the external perimeter of 2D random curves at scaling limit, which is predicted by the Schramm-Loewner evolution (SLE). 
\end{abstract}
\maketitle

{\it Introduction.} 
In polymer solution beyond the overlap concentration, 
adaptation of the polymer configurations in 2D is dramatically different from that in 3D. 
At thermodynamic equilibrium, polymer chains in strict 2D are bound to segregate and become territorial, forming entanglement-free polymer domains, whereas the chains in 3D tend to interpenetrate and are entangled to maximize the entropy of polymer solution~\cite{deGennesbook}.  
{\color{black}
For 2D polymer solution, multilayered 2D polymer solution with finite thickness, called ``self-avoiding trails''~\cite{meirovitch1989surface,semenov2003EPJE}, in which polymer chains overlay on top allowing chain intersections on a projected 2D plane, may physically be more realistic and relevant as it represents thin polymer films, polymers at interfaces or under nano-confinement~\cite{luengo1997thin,jones1999dynamics,katsumata2018glass}. 
Yet, there have also been a number of studies exploring transitions in the physical property of 2D polymer solution with decreasing thickness of the solution, from the one with more entanglement-rich 3D bulk-like property to the other with entanglement-free 2D surface-like property~\cite{frank1996polymer,bay2018confinement,Liu19Nanolett,pressly2019increased}.   
While there is a limited number of experimental investigation on 2D polymer monolyers, their physical properties are expected to play significant roles in many practical applications~\cite{langevin2010interfacial}. 
Furthermore, the system of many polymer chains confined in strict 2D, 
especially in \emph{dense} phase, has been an abiding theoretical interest in polymer physics~\cite{duplantier1986JPA,saleur1987PRL,duplantier1987critical,duplantier1987exact,duplantier1989PR,duplantier1987PRB,Duplantier89JSP,Duplantier87PRL,carmesin1990JP,semenov2003EPJE,meyer2009perimeter,meyer2010static,schulmann2013PSSC}.   

Here, we perform numerical simulations of strict 2D polymer solutions consisting of polymer chains under two distinct solvent conditions, i.e., good and $\Theta$ solvents which can be realized by tuning the strength of inter-monomer interaction~\cite{Liu2022JPCB}, at varying area fraction ($\phi$), and study the $\phi$-dependent variation in the geometry of polymer domain boundaries.}
The outer boundaries of the domains in 2D are not smooth but irregular 
\cite{semenov2003EPJE,meyer2009perimeter,meyer2010static,schulmann2013PSSC}, and the extent of the irregularity 
can be quantified using the fractal dimension~\cite{mandelbrot1982fractal}. 
Specifically, the ``external perimeter'' ($E_p$) 
increases with ``the {\color{black}root-mean-squared} size of monomers constituting the perimeter,'' $R_p\equiv \left(\frac{1}{2N_p^2}\sum_{i,j\in\partial}^{N_p}(\vec{r}_i-\vec{r}_j)^2\right)^{1/2}$ with $\vec{r}_i$ the position of $i$-th monomer and $N_p$ the number of monomers making up the external perimeter ($\partial$), and 
{\color{black}the fractal (Hausdorff) dimension of the external perimeter~\cite{mandelbrot1982fractal,semenov2003EPJE,meyer2009perimeter,meyer2010static,schulmann2013PSSC}, $\mathcal{D}^{\partial}\in[1,2]$, can be obtained from the scaling relation between the two quantities averaged over many polymer configurations 
\begin{align}
\langle E_p\rangle\sim \langle R_p\rangle^{\mathcal{D}^{\partial}}, 
\label{eqn:fractal}
\end{align} 
where $\langle\ldots\rangle$ denotes the ensemble average.}

Calculating $\mathcal{D}^{\partial}$ through the numerics of monodisperse polymer solutions with varying $\phi$, we discover that $\mathcal{D}^{\partial}$ exhibits qualitatively different $\phi$-dependences on the solvent quality. 
Remarkably, $\mathcal{D}^{\partial}$ in good solvents ($\mathcal{D}_{\rm SAW}^\partial$) exhibits a non-monotonic variation with $\phi$, whereas $\mathcal{D}^{\partial}$ of $\Theta$ chains ($\mathcal{D}_\Theta^\partial$) remains constant over the same range of $\phi$.  
The constant $\mathcal{D}^{\partial}_\Theta$ can be understood as an outcome of the compensation between attraction and repulsion that characterizes the nature of $\Theta$ chain \cite{coniglio1987PRB,Duplantier87JCP,Liu19Nanolett,Jung2020Macromol,Liu2022JPCB}. 
Despite a number of works that analyzed the irregularity of polymer domain boundary in 2D polymer solution \cite{saleur1987PRL,semenov2003EPJE,beckrich2007macro,meyer2009perimeter,meyer2010static,schulmann2013PSSC}, {\color{black}these studies are mainly focused on polymer melts or dense polymer systems.} 
Our finding of the non-monotonic variation of $\mathcal{D}^{\partial}_{\rm SAW}(\phi)$, especially over  intermediate concentrations, has not been reported elsewhere.

Since long polymers {\color{black}($N\gg 1$)} are considered as an critical object in scaling limits (see {\bf Polymer-magnet analogy and critical exponents} in the Appendix), 
the fractal dimension of polymer as well as the values of $\mathcal{D}^{\partial}$ can be associated with the fundamental scaling exponents. 
Here we first investigate the origin of $\phi$-dependent variation of $\mathcal{D}^{\partial}$ 
in the language of polymer in 2D. We also examine the problem under the hood of 
Schramm-Loewner evolution (SLE), an elegant mathematical tool that utilizes the properties of conformal invariance to offer quantitative description for the boundaries of 2D critical systems at their scaling limits~\cite{kager2004guide,rohde2005basic,cardy2005sle,lawler2009conformal}.

{\color{black}
\section{Methods}
\subsection*{Generating polymer solution in two dimensions.}  
In order to simulate a single polymer chain composed of $N$ segments under two distinct solvent conditions, we used the energy potential, 
$\mathcal{H}({\bf r}) = \mathcal{H}_{b}({\bf r}) + \mathcal{H}_{nb}({\bf r})$, 
where ${\bf r} = \{ {\bf r}_{i} \}$ and ${\bf r}_{i}$ is the coordinate of the $i$-th monomer ($i=1,2,\ldots,N-1$) in a 2D surface. 
The term $\mathcal{H}_{b}({\bf r})$ models the chain connectivity using the finite extensible nonlinear elastic (FENE) potential and a shifted Weeks-Chandler-Anderson (WCA) potential, 
\begin{widetext}
\begin{align}
\label{hb}
\mathcal{H}_{b}({\bf r}) &= -\frac{k}{2} R^{2}_{c} \sum_{i=0}^{N-1} \log\left(1-\frac{r_{i,i+1}^{2}}{R^2_c}\right)
 + \sum_{i=1}^{N-1} 4\varepsilon_{\rm SAW}\left[\left(\frac{a}{r_{i,i+1}}\right)^{12} - \left(\frac{a}{r_{i,i+1}}\right)^{6} + \frac{1}{4}\right] H(2^{1/6} - r_{i,i+1}/a), 
\end{align}
\end{widetext}
where $r_{i,i+1} \equiv |{\bf r}_{i+1} -{\bf r}_{i}|$ is the segment length, $H(\cdots)$ is the Heaviside step function. 
The energy potential with the parameters $ k=30k_BT/a^2$, $R_{c} = 1.5a$, and $\varepsilon_{\rm SAW}/k_BT=1$ equilibrates the segments at around the van der Waals distance between monomers ($a$). 
The non-bonded interactions between two monomers ($|i-j|\geq 2$) are represented by the term $\mathcal{H}_{nb}({\bf r})$.  
To generate polymer chains in two distinct solvent qualities, 
we used different expressions of $\mathcal{H}_{nb}({\bf r})$. 
For polymers in good solvent, $\mathcal{H}_{nb}({\bf r})=\mathcal{H}_{nb}^\text{good}({\bf r})$, 
\begin{widetext}
\begin{align}
\mathcal{H}_{nb}^\text{good}({\bf r}) = 
  \sum_{i<j} 4\varepsilon_{\rm SAW}\left[\left(\frac{a}{r_{ij}}\right)^{12} - \left(\frac{a}{r_{ij}}\right)^{6} 
+ \frac{1}{4}\right] H(2^{1/6} - r_{ij}/a). 
\label{eqn:Hnb_SAW}
  \end{align}
  For polymers in $\Theta$ solvent, 
  $\mathcal{H}_{nb}({\bf r})=\mathcal{H}_{nb}^\Theta({\bf r})$
  \begin{align}
\mathcal{H}_{nb}^{\Theta}({\bf r}) =    \sum_{i<j} \varepsilon_\Theta\left[\left(\frac{a}{r_{ij}}\right)^{12} - 2\left(\frac{a}{r_{ij}}\right)^{6} + \Delta_s\right] H(2.5 - r_{ij}/a), 
\label{eqn:Hnb_Theta}
\end{align}
\end{widetext}
%
where we set $\varepsilon_{\Theta}/k_BT = 1.013$ and $\Delta_{s} = 2\times(4/10)^{6} - (4/10)^{12}$, the latter of which is introduced to connect the potential at $r_{ij}=2.5a$ smoothly to $\mathcal{H}_{nb}^{\Theta}({\bf r})=0$ for $r_{ij}\geq 2.5a$. 
As shown in our previous study\cite{Liu2022JPCB}, 
the energy potentials with Eq.\ref{eqn:Hnb_SAW} and Eq.\ref{eqn:Hnb_Theta}, corresponding to the good and $\Theta$ solvent conditions, generate polymer chains whose root mean squared end-to-end distance ($R_{ee}\equiv \langle R_{ee}^2\rangle^{1/2}$) scales as $R_{ee}\sim N^{3/4}$ and $R_{ee}\sim N^{4/7}$, respectively. 

For efficient sampling of polymer configruations at equilibrium, 
we integrated the equation of motion of polymer chain coupled to a Langevin thermostat in underdamped regime:
$m {\ddot {\bf r}}_{i} = -\zeta {\dot {\bf r}}_{i} - \nabla_{{\bf r}_i} \mathcal{H}({\bf r}) + {\bf \xi}_{i}(t)$,
with the random force satisfying $\langle {\bf \xi}_{i}(t)\rangle=0$ and $\langle{\bf \xi}_{i}(t) \cdot {\bf \xi}_{j}(t') \rangle=4 \zeta k_{B}T \delta_{ij}\delta(t-t')$. 
A small time step $\delta t = 0.005 \tau$ and a friction coefficient $\zeta = 0.1 m/\tau$, which yield the characteristic time scale $\tau = \left(m a^{2}/\varepsilon\right)^{1/2}$ with $\varepsilon\simeq \varepsilon_{\rm SAW}$ or $\varepsilon_\Theta$, were employed.

For the solutions of monodisperse polymer chains with varying lengths ($N=40$, 80, 160, 320, and 640), 
inter-chain monomer interactions were identically modeled as those for intra-chain monomers (Eqs.~\ref{eqn:Hnb_SAW} and ~\ref{eqn:Hnb_Theta}). 
The polymer solution with varying $\phi$ was simulated in two steps. 
(i) From a condition of dilute phase ($\phi\approx 7.85\times 10^{-3}$) that contains $36$ pre-equilibrated chains, 
the size of the periodic box was reduced step by step with $L\rightarrow \eta L$ ($\eta = 0.904$), so that 
the area fraction is increased by a factor of $\eta^{-2}$ in each step.  
At each value of $\phi$, overlaps between monomers, induced by an excessive shrinkage of the box, were eliminated by gradually increasing the short-range repulsion part of $\mathcal{H}$. 
More specifically, the non-bonded potential $\mathcal{H}_{nb}({\bf r})$ was replaced with $\min\{u_{c}, \mathcal{H}_{nb}({\bf r})\}$, in which $u_{c}$ was slowly elevated. 
(ii) For the production run, the system was simulated for $500N\tau$, and chain configurations were collected every $0.1N\tau$. 
For each combination of $N$ and $\phi$, 10 replicas were generated from different initial chain configurations and random seeds. 
The simulations were performed using the ESPResSo 3.3.1 package \cite{limbach2006espresso}.

\subsection{External perimeter of polymer domain and fjord-like configurations}
The external perimeter of a given polymer configuration ($E_p$), defined as the length of closed path on a square lattice with the lattice spacing $l$, was calculated by employing the turn-right tie-breaking rule~\cite{saberi2009thermal,Saberi2011PRE} and its average value ($\langle E_p\rangle$) as well as the average gyration radius ($\langle R_p\rangle$) was obtained over thousands of polymer configurations.  
Since the polymer chain are simulated in continuous space, the absolute size of $\langle E_p\rangle$ is altered by the lattice spacing $l$; yet, the fractal dimension, corresponding to the scaling exponent defined 
between $\langle E_p\rangle$ and $\langle R_p\rangle$ is insensitive to the $l$ as long as $l$ is sufficient small compared to other length scales, i.e., $l\ll \langle E_p\rangle, \langle R_p\rangle$. 
For the analysis in this study, we used $l=a$. 

Next, the ``fjord"-like configuration along the perimeter discussed in this study is identified as the ``closed segment" where two segments of the perimeter on the square lattice meet each other at a single lattice point.  
Small fjords were removed from the list if they were part of bigger fjords. 
}

\begin{figure*}[ht!]
\includegraphics[width=0.88\textwidth]{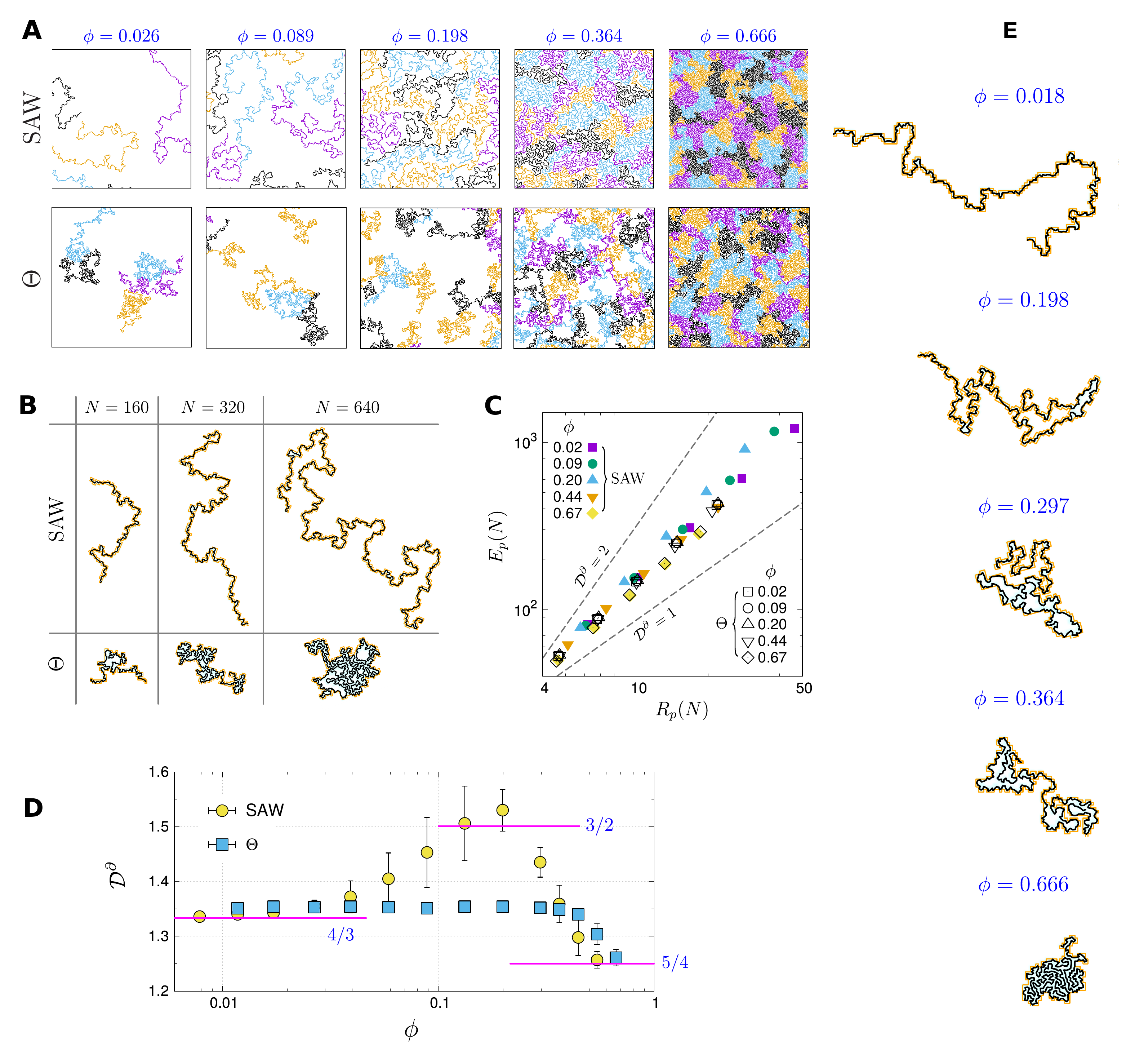}
\caption{Fractal dimension of polymer domain boundary with varying area fraction $\phi$. 
(A) Polymer solutions of SAW (top) and $\Theta$ chains with $N=640$ in 2D with increasing $\phi$ (see Ref.~\cite{Liu2022JPCB} for the details of simulations).   
Each panel{\color{black}, visualizing the polymer solution consisting of strictly non-crossing polymer chains,} is drawn in the 2D box of the identical size. 
(B) Configurations of SAW and $\Theta$ chains with varying $N$ in dilute solution ($\phi<\phi^\ast$). 
The overlap area fraction ($\phi=\phi^\ast$), in which the intra-monomer concentration (area fraction, $\phi$) is comparable to the inter-monomer concentration, i.e., $\phi^\ast\sim Na^d/R_F^d\sim N^{1-\nu d}$, has previously been determined at $\phi^\ast\approx 0.018$ and $\phi^\ast\approx 0.266$ for the polymer solution with $N=640$ under a good and $\Theta$ solvent condition, respectively \cite{Liu2022JPCB}.  
External perimeter (orange) of a polymer configuration (black) is calculated based on the turn-right tie-breaking rule \cite{saberi2009thermal,Saberi2011PRE}. 
The interior of the domain enclosed by the perimeter is colored in pale blue. 
(C) Log-log plot of the external perimeter of the chain ($E_p$) versus the gyration radius of the perimeter ($R_p$)  are produced using the chains with five different lengths ($N=40$, 80, 160, 320, and 640) for a given value of $\phi$. 
(D) The fractal dimension $\mathcal{D}^{\partial}$ was calculated from 
the data points in (C). 
The three characteristic values of $\mathcal{D}^{\partial}=4/3$, 3/2, and 5/4, are marked in blue. 
(E) Typical configurations of SAW at varying $\phi$'s.}
\label{perimeter}
\end{figure*}

\section{Results and Discussions} 
\subsection{Rationalizing the values of $\mathcal{D}^{\partial}$ using polymer arguments} 
Simulating 2D monodisperse polymer solution (see Methods~\cite{Liu2022JPCB} and Fig.~\ref{perimeter}), 
we study the configurations of polymer domains. 
To examine the $\phi$-dependent irregularity of the domain boundary (Fig.~\ref{perimeter}A),  
we calculate {\color{black}$\langle E_p\rangle$} and analyze its variation against {\color{black}$\langle R_p\rangle$}, and extracted $\mathcal{D}^{\partial}$ as defined in Eq.\ref{eqn:fractal} {\color{black}(see Fig.~\ref{perimeter}C. Note that including Fig.~\ref{perimeter}C, we hereafter use simplified notation $E_p$ and $R_p$ without $\langle\ldots\rangle$ denoting the ensemble average over many polymer configurations)}. 
As shown in Fig.~\ref{perimeter}D, in good solvents, 
$\mathcal{D}^{\partial}$ of polymer chains (SAWs) exhibits a non-monotonic variation with $\phi$ (Fig.~\ref{perimeter}D), starting from $\mathcal{D}^{\partial}_{\rm SAW}=4/3(\approx 1.33)$ in dilute solution ($\phi\approx 0$), maximizing to $\approx 3/2$ at an area fraction ($\phi=\phi_
{cr}\simeq 0.2$) {\color{black}corresponding to a cross-over point,} and reaching $5/4(=1.25)$ in a dense phase ($\phi\approx 1$). 
On the other hand, $\mathcal{D}^{\partial}$ of chains in $\Theta$ solvents remains constant ($\mathcal{D}^{\partial}_\Theta\simeq 4/3$) over the range of $\phi$ up to $\phi\lesssim 0.4$, and also drops to $5/4 (= 1.25)$ at $\phi\approx0.67$. 
Note that our calculation of the fractal dimension of external perimeter confirms the Mandelbrot conjecture \cite{Lawler2000TheDO}, namely, $\mathcal{D}^{\partial}=\mathcal{D}^{\partial}_{\rm B}=4/3$ for the frontier (outer boundary) of the traces generated from planar Brownian motion (see {\bf Hausdorff dimension of planar Brownian frontier}
in the Appendix and Fig.~\ref{planarBrown}). 
The values of $\mathcal{D}^{\partial}$ in Fig.~\ref{perimeter}D under the limiting and the crossover conditions are rationalized below.

(i) For 2D SAW chain in dilute phase ($\phi<\phi^\ast$), the majority of the monomers are exposed to the solvent (see the configurations of SAW in Fig.~\ref{perimeter}B). 
As a result, the perimeter of the polymer domain is expected to scale with the number of monomers, $E_p(N)\sim N$. 
Just like the mean squared size of polymer and the mean squared end-to-end distance obey the same scaling relation $R^2_F\sim \langle R^2_{ee}\rangle \sim N^{2\nu}$, $R_p$ displays a scaling of $R_p\sim N^\nu$ with $\nu=\nu_{\rm SAW}=3/4$. 
Hence, along with Eq.\ref{eqn:fractal} it is expected that $E_p\sim N^{\nu \mathcal{D}^{\partial}}\sim N$; therefore, $\mathcal{D}^{\partial}_{\rm SAW}=\nu^{-1}_{\rm SAW}=4/3\approx 1.33$~\cite{schulmann2013PSSC}.

(ii) Configurations of $\Theta$ chain differ from those of SAW (Fig.~\ref{perimeter}B) in that 
some monomers are buried inside the domain, whereas others are exposed to the periphery constituting the external perimeter.
$\Theta$ chains in dilute phase obey $N\sim R_F^{\mathcal{D}_{\Theta}}$, characterized by the fractal dimension of percolating clusters $\mathcal{D}_{\Theta}=7/4$~\cite{coniglio1987PRB,Sapoval85JPL,Bunde85JPA,Duplantier87PRL,saleur1987PRL,duplantier1989PRL,duplantier1989PR,Liu19Nanolett,Jung2020Macromol}; however, {\color{black}Figure~\ref{perimeter}D} indicates that the external perimeter of the $\Theta$ chain is still self-avoiding, such that 
$\mathcal{D}_{\Theta}^\partial=\mathcal{D}_{\rm SAW}=4/3$~\cite{grossman1986structure,grossman1987accessible,aizenman1999PRL}. 

\begin{figure}[t]
\includegraphics[width=0.5\textwidth]{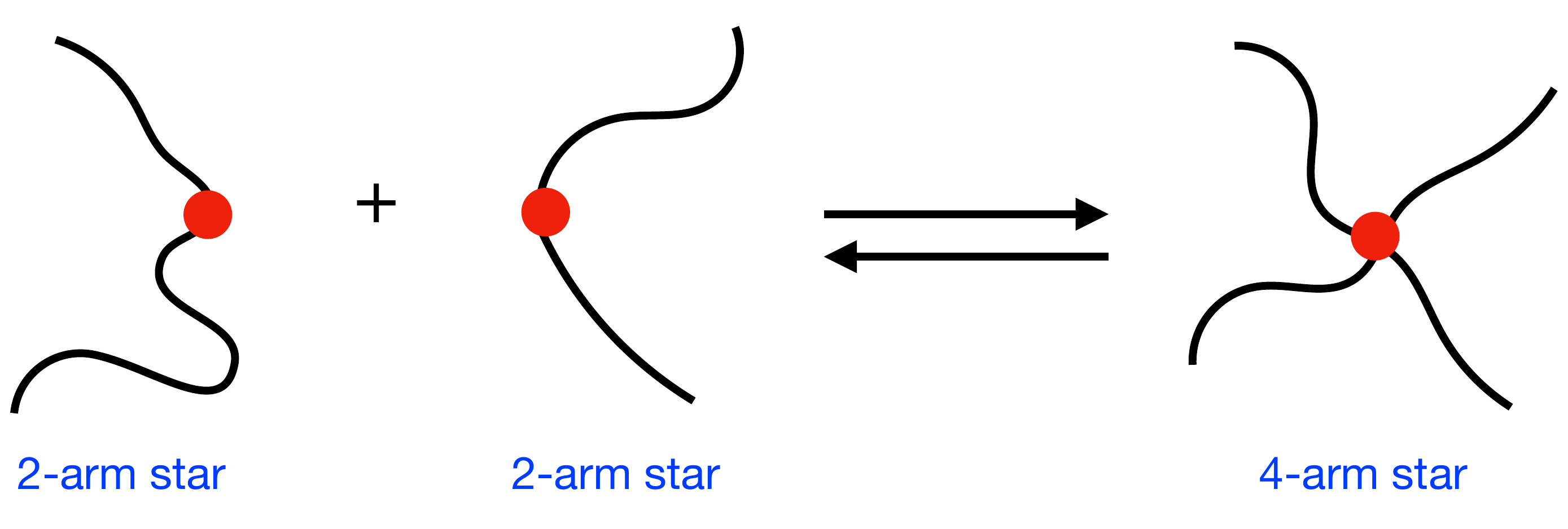}
\caption{{\color{black}Illustration of 4-arm star polymer made of two 2-arm star polymers. 
For the total numbers of configurations (partition sums) for 4-arm star polymer and 2-arm star polymer 
given as $Z_{N,4}\sim \mu^{4N}N^{\gamma_4-1}$ and $Z_{N,2}\sim \mu^{2N}N^{\gamma_2-1}$, 
the inter-polymer contact probability $f_{\rm int}$ is expected to scale as $f_{\rm int}\sim Z_{N,4}/Z_{N,2}^2$. 
}}
\label{partitionsum}
\end{figure}

(iii) In dense polymer solution, the external perimeter of a polymer domain 
is proportional to the number of interchain contacts \cite{schulmann2013PSSC}. 
\begin{align}
E_p(N)\propto N\times  f_{\rm inter}\sim N\frac{Z_{N,4}}{Z^2_{N,2}}
\label{eq:Ep}
\end{align}
where $f_{\rm inter}$, the fraction of such contacts per chain, can be associated with 
{\color{black}the ratio of total numbers of polymer configurations between the 4-arm star polymer and the product of two 2-arm star polymers (see Figure~\ref{partitionsum}), in other words, it is the ratio of}
the partition sums between $4$-arm ($Z_{N,4}$) and two 2-arm star polymers ($Z^2_{N,2}$), i.e., 
 $f_{\rm inter}\sim Z_{N,4}/Z^2_{N,2}$. 
As the partition sum of an $L$-arm star polymer with each arm consisting of $N$ segments is asymptotically ($N\gg 1$) related with $\mu$ the connectivity constant and $\gamma_L$ the enhancement exponent as $Z_{N,L}\sim \mu^{LN}N^{\gamma_L-1}$\cite{duplantier1986PRL,Duplantier89JSP}, we obtain $f_{\rm inter}=N^{\gamma_{4}-2\gamma_2+1}$ and $E_p(N)\sim N^{\gamma_{4}-2\gamma_2+2}$~\cite{semenov2003EPJE}. 
From $E_p(N)\sim N^{\nu \mathcal{D}^{\partial}_{\rm D}}$, it follows that 
\begin{align}
\mathcal{D}^{\partial}_{\rm D}&=\frac{1}{\nu}(\gamma_4-2\gamma_2+2). 
\label{DD}
\end{align} 
Using the concepts of polymer-magnet analogy, Duplantier derived the exact expression of the enhancement exponent of 2D L-arm star polymer as 
$\gamma_L=9/8+(3-L)L/32$ (see Eq.~\ref{eqn:enhancement_D} and the Appendix for details of derivation)~\cite{duplantier1989PR,Duplantier89JSP}. 
Since $\gamma_2=19/16$, $\gamma_4=1$, and $\nu=1/2$ for polymers in dense phases, we obtain $\mathcal{D}^\partial_{\rm D}=5/4$.

(iv) For polymer chains in the polymer solution in good solvents, 
the fractal dimension of the domain boundary changes non-monotonically with $\phi$ and maximizes to $\mathcal{D}^{\partial}_{\rm SAW}\approx 1.5$ at $\phi_{cr}\approx 0.2$ (Fig.~\ref{perimeter}D).  
According to the polymer configurations visualized at the five different values of $\phi$ in Fig.~\ref{perimeter}E, 
the outer boundary of the chain at dilute phase ($\phi \approx 0.018$) is made of essentially all the monomers constituting a polymer chain.  
However, as the $\phi$ increases and the overall size of polymer domain decreases~\cite{Liu2022JPCB}, 
some of the monomers are engulfed inside the domain, which divides the polymer domain into the interior (colored in pale blue) and the exterior. 
In particular, the monomers constituting the exterior of the domain are used to define the external perimeter (a closed loop depicted with an orange line).  
Remarkably, it appears that the ruggedness (or fractal dimension) of this external perimeter also changes non-monotonically with $\phi$ and maximizes at an intermediate value of $\phi$ ($\phi=0.198$ and $0.297$), and it flattens out at the highest value ($\phi=0.666$). 

\begin{figure*}[t]
\includegraphics[width=0.75\linewidth]{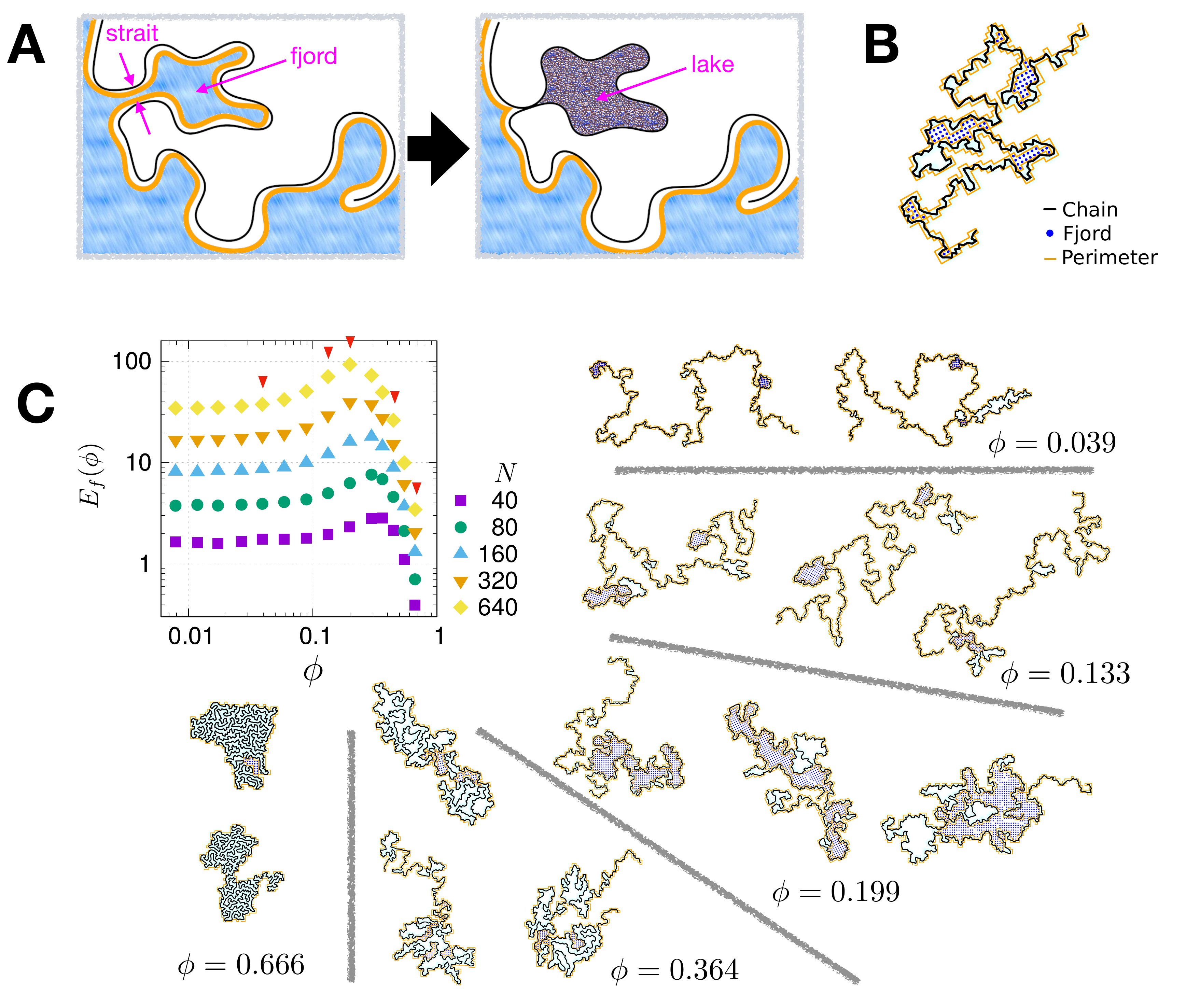}
\caption{Contribution of fjords to the perimeter ruggedness. 
(A) An illustration of a fjord (left) transforming into a lake (right). 
When a narrow strait merges, the fjord turns into a lake, which smoothes out the perimeter (orange) of domain boundary. 
(B) A polymer configurations (black line) with fjords, which are defined as the \emph{closed segments} of the external perimeter (orange line) {\color{black}calculated in the unit of the lattice spacing $a$} on a square lattice.  
The region surrounded by the fjords is demarcated using the blue dots.  
(C) The average length of fjords per chain ($E_f$) with increasing $\phi$.  
The maximal $E_f$ is identified at $\phi=(0.2-0.3)$, gradually shifting towards the smaller $\phi$ as $N$ increases. 
Depicted next the graph are polymer configurations (black line) with $N = 640$ at varying $\phi$'s ($\phi=0.039$, 0.133, 0.199, 0.364, and 0.666), which are also marked with the red arrows on the graph.  
}
\label{fjords}
\end{figure*}

A careful visual inspection of polymer configurations suggests that 
the outer boundary of polymer domain are characterized with ``fjord"-like configurations with narrow ``straits"~\cite{grossman1986structure,grossman1987accessible,aizenman1999PRL}, the area of which increases up to some $\phi$. 
When $\phi$ further increases and the solution is in the dense phase, 
the merging straits transform those fjords into lakes, which smoothes out the domain boundary (see Fig.~\ref{fjords}A for the illustration of a fjord-like configuration in the domain boundary, a strait, and a lake).  
More specifically, as shown in Fig.~\ref{fjords}B, 
for a given polymer configuration (black curve), 
the external perimeter of the polymer domain (orange line) is identified on a 2D square lattice by employing the turn-right tie-breaking rule~\cite{saberi2009thermal,Saberi2011PRE}.
In Fig.~\ref{fjords}B, the ``fjord"-like configuration  
corresponds to the part of perimeter forming a closed loop, and the region enclosed by the loop is marked with the blue dots highlighting the areas occupied by the fjords formed along the domain boundary.   
We find that the total length of such fjords that contributes to the irregularity of the domain boundaries is short {\color{black}in dilute phase ($\phi<\phi^\ast$, where $\phi^\ast$ denotes the overlap concentration)}, and it gradually increases up to $\phi\approx (0.2-0.4)${\color{black}, which is significantly greater than $\phi^\ast$,} and decreases at higher $\phi$.   
Explicit calculation of the average contour length of the fjords per chain exhibits non-monotonic variations (Fig.~\ref{fjords}C), which is similar to that of {\color{black}the $\phi$-dependent fractal dimension of domain boundary ($\mathcal{D}^\partial(\phi)$)} shown in Fig.~\ref{perimeter}D.  
We surmise that the gradual increase of osmotic pressure~\cite{Liu2022JPCB} exerted by the neighboring chains facilitates the folding of intra-domain boundary to shape polymer configurations with rugged perimeter, reminiscent of fjords, until they are engulfed into the interior of the domain at dense phases.

\subsection{Schramm-Loewner evolution}
{\color{black}
The Schramm-Loewner evolution (SLE), which uses the property of conformal invariance, 
is an elegant mathematical apparatus developed in early 2000s~\cite{kager2004guide,rohde2005basic,lawler2009conformal}. 
It was conjectured to describe any critical statistical mechanical object in the form of non-crossing stochastic paths in 2D using a one-dimensional Brownian motion~\cite{kager2004guide,rohde2005basic,lawler2009conformal}, offering an entirely different perspective to understand a number of issues in 2D critical phenomena~\cite{cardy2005sle}. 
Assuming that the outer boundary of polymer domain in 2D is a conformally invariant geometrical object, 
we consult the SLE to cross-check and better understand our findings of the $\phi$-dependent 2D curve of polymer domain boundary.} 

\begin{figure*}[t]
\includegraphics[width=1.0\linewidth]{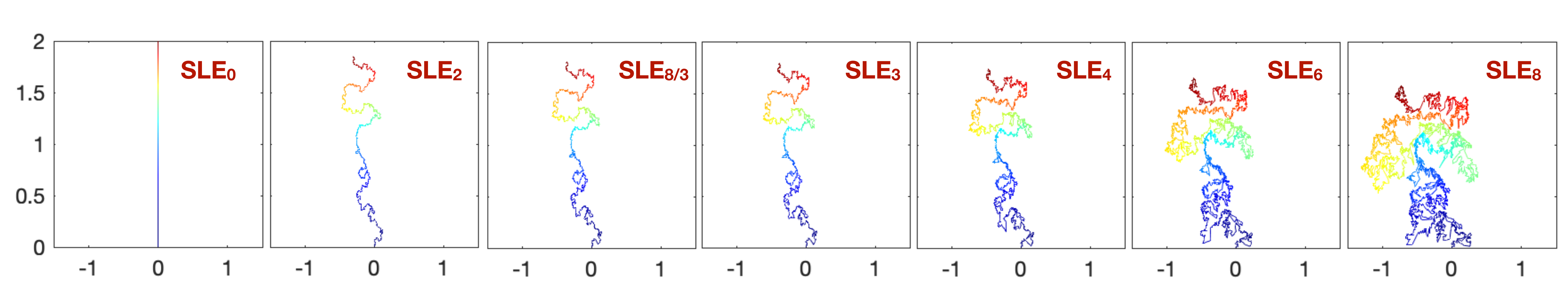}
\caption{{\color{black}The 2D random curves $\gamma(t)=g^{-1}_t(a_t)$, i.e., $SLE_\kappa$, generated using the inverse mapping of the conformal map $g_t(z)$ satisfying the differential equation given in Eq.\ref{eqn:SLE} for various values of $\kappa$~\cite{kennedy2009numerical}. 
All the curves grow in time from dark blue ($t=0$) to dark red ($t=10^4$).  
$SLE_\kappa$ with $\kappa=0$, 2, 8/3, 3, 4, 6, and 8 correspond to the growing stick,  
loop-erased random walk, SAW, domain wall of 2D Ising model at the critical point, Gaussian free field, cluster boundaries in percolation, and space-filling curve, respectively~\cite{cardy2005sle}.}
}
\label{sle}
\end{figure*}

Specifically, {\color{black}SLE is based on} a conformal map $w=g_t(z)$ that satisfies the following differential equation 
\begin{align}
\partial_tg_t(z)=\frac{2}{g_t(z)-a_t}
\label{eqn:SLE}
\end{align}  
with $g_0(z)=z$ for $z\rightarrow \infty$.   
{\color{black}The map $g_t(z)$} uniformizes 2D curves, $\gamma[0,t]$, in the upper half-plane ($\mathbb{H}=\{{\rm Im}(z)>0;z\in \mathbb{C}\}$) onto the real axis such that {\color{black}a 2D curve at time $t$, $\gamma(t)$, is mapped onto the real value $a_t$ in the real axis via the conformal map,} $a_t=g_t(\gamma(t))\in \mathbb{R}$~\cite{kennedy2009numerical}.  
If {\color{black}the $a_t$, called a driving function, is deterministic with $\kappa=0$, the resulting 2D curve is a growing stick ($SLE_0$, the left-most panel in Figure~\ref{sle}), which is the solution of the Loewner's differential equation introduced in 1923~\cite{lowner1923untersuchungen}.   
On the other hand, if the $a_t$} is stochastic, 
satisfying $\langle a_t\rangle =0$ and $\langle a_t^2\rangle =\kappa t$ with 
$\kappa$ corresponding to the diffusivity of the Brownian motion, 
the inverse mapping $g_t^{-1}(\omega)$, namely $\gamma(t)=g_t^{-1}(a_t)$ can generate  
{\color{black}a 2D non-crossing random curve, called $SLE_{\kappa}$ visualized in Figure~\ref{sle} for some physically relevant values of $\kappa$,} whose behavior is decided solely by the value of a single parameter $\kappa$. 

The fractal (Hausdorff) dimension of the $SLE_{\kappa}$ is given as \cite{beffara2008dimension,rohde2005basic} 
\begin{align}
\mathcal{D}=\min{\left(2,1+\kappa/8\right)}. 
\label{eqn:Df}
\end{align}
Further, it has been conjectured that the outer boundary of $SLE_\kappa$ for $\kappa\geq 4$ corresponds to the curve of an $SLE_{16/\kappa}$, which is known as ``SLE duality'' \cite{duplantier2000conformally,beffara2008dimension}.  
Thus, for $\kappa \geq 4$, the fractal dimension of the outer boundary is   
\begin{align}
\mathcal{D}^{\partial}=1+2/\kappa. 
\label{eqn:Dp}
\end{align}
Eq.~\ref{eqn:Dp} can be used to validate our results in Fig.~\ref{perimeter}D. 

(i) The fractal dimension of $\Theta$ chain ($\kappa=6$) is $\mathcal{D}_\Theta=7/4$ (Eq.~\ref{eqn:Df}), and that of its outer boundary is $\mathcal{D}^{\partial}_{\Theta}=4/3$ (Eq.~\ref{eqn:Dp}). 
The exponents obtained from our numerics in Fig.~\ref{perimeter}D are in perfect agreement with the values predicted by the SLE. 
Notably, despite the disparate polymer configurations in the two solvent conditions (see Fig.~\ref{perimeter}B), 
{\color{black}the fractal dimension of polymer domain boundary of $\Theta$ chain ($\mathcal{D}^\partial_{\Theta}$) is still identical to that of SAW, namely, $\mathcal{D}^\partial_{\Theta}=\mathcal{D}_{\rm SAW}$.} 
In fact, the external perimeter of the ideal random planar walk (Brownian motion)
is also self-avoiding (Fig.~\ref{planarBrown}), which is known as the Mandelbrot's conjecture~\cite{Lawler2000TheDO}.  

(ii) From the perspective SLE, polymer chain in dense solution is space-filling with the corresponding value of $\kappa$ being $\kappa\geq  8$~\cite{rohde2005basic,kager2004guide,cardy2005sle}. 
In this case, we get $\mathcal{D}=\mathcal{D}_{\rm D}=2$ (Eq.\ref{eqn:Df}) and $\mathcal{D}^{\partial}=\mathcal{D}^{\partial}_{\rm D}=5/4$ (Eq.\ref{eqn:Dp}). 

(iii) Lastly, SLE can be used to account for the $\phi$-dependent non-monotonic variation of $\mathcal{D}^\partial_{\rm SAW}$, the origin of which we have ascribed to the average length of fjords per chain that also exhibits a non-monotonic variation.   
A transformation $h_t(z)=g_t(z)-a_t$ casts Eq.\ref{eqn:SLE} into 
$\partial_th_t(z)=2/h_t(z)+\xi_t$ where $\xi_t$ is a white noise satisfying $\langle \xi_t\rangle=0$ and $\langle \xi_t\xi_{s}\rangle=\kappa\delta(t-s)$.  
If one considers the dynamics of SLE curves projected on the real axis $\Re{z}=x\in \mathbb{R}$, $h_t(x)=x_t$, 
the $x_t$ is described by the 1D stochastic dynamics, known as \emph{the Bessel process}~\cite{bauer20062d,cardy2005sle}
\begin{align}
\frac{dx_t}{dt}=\frac{2}{x_t}+\xi_t. 
\end{align}
Heuristically, 
the nature of the dynamics $x_t$ is dictated by the value of $\kappa$, and a crossover between a deterministic ($x_t^2\sim 4t$) and a stochastic growth ($\langle x_t^2\rangle\sim \kappa t$) occurs at $\kappa=4$. 
{\color{black}Along with the SLE curves depicted with varying $\kappa$ in Figure~\ref{sle},} this means that for $\kappa<4$ the unprojected, original SLE curves, $\gamma[0,t]$, are simple {\color{black}and more deterministic} and neither hit the real axis nor have self-intersections, whereas they {\color{black}become more stochastic and} self-intersecting for $\kappa> 4$~\cite{kager2004guide,rohde2005basic,cardy2005sle}. 
The SLE curves at $\kappa=4$ correspond to those lying precisely at the crossover point between the two contrasting behaviors.  
According to the SLE duality (Eq.\ref{eqn:Dp}), the fractal dimension of the external perimeter of SLE curves is upper bounded by $\mathcal{D}^\partial_{\rm max}=3/2$ for $\kappa=4$, and this number is consistent with the maximal value of $\mathcal{D}^\partial$ calculated in Fig.~\ref{perimeter}D, i.e., $\mathcal{D}^\partial_{\rm max}\approx 1.53\pm 0.04$.

\section{Conclusion}
In summary, examining the $\phi$-dependent ruggedness of the polymer domain boundary of individual polymer chain in polymer solution ($\mathcal{D}^\partial(\phi)$), we discover the non-monotonic variation in the ruggedness for 2D polymer solution in good solvents. 
The values of $\mathcal{D}^\partial$ at the crossover point as well as under the limiting conditions are rationalized using the fundamental critical exponents of polymer configurations in 2D ($\nu$, $\gamma_2$, and $\gamma_4$) and the idea of SLE.   
Among them, of particular note is the maximal ruggedness $(\mathcal{D}^\partial_{\rm SAW})_{\rm max}=3/2$, which the SLE (Eq.\ref{eqn:Dp}) ensures as the universal upper bound of the fractal dimension for 2D interface.  
Interestingly, similar values of maximal fractal dimension have been reported for the fractal interface in bacterial biofilms~\cite{qin2021hierarchical,brooks2022computational}.
We have associated $(\mathcal{D}^\partial_{\rm SAW})_{\rm max}$ with the maximal ``fjord"-like corrugations that result from the marginal folding of domain boundary (see Fig.~\ref{fjords}A).  
The adaptation of polymer configurations with $\phi$, which gives rise to the non-monotonic variation of irregularity in the domain boundary, is unique and fundamentally differs from the $\phi$-dependent variation of osmotic pressure ($\Pi$) in that the latter is dictated by the exponent $\nu$ alone and displays monotonic variation with $\phi$
~\cite{vilanove1980PRL,witte2010macromolecules,goicochea2015scaling,Liu2022JPCB}. 

Finally, our finding of the $\phi$-dependent corrugations in polymer domain boundary  is amenable to experimental verification which may require either a direct/indirect visualization of polymer configurations~\cite{jones1999chain,kumaki2010visualization,sugihara2012visualization} or a careful investigation on rheological responses of ultrathin polymer films {\color{black}or polymer monolayers at air-water interface~\cite{o2005rheological,langevin2010interfacial}. 
In light of our finding that the ruggedness of domain boundary changes non-monotonically and maximizes at a cross-over area fraction ($\phi_{\rm cr}\approx 0.2$) 
that lies between $\phi^\ast$ and $\phi\lesssim 1$, it is possible that a certain dynamical behavior, such as the amoeba-like fluctuations of the (sub)chain contours with a relaxation dynamics of $\tau\sim N^{15/8}$ conjectured~\cite{semenov2003EPJE} and explicitly observed in a simulation study~\cite{wittmer2010algebraic} for a polymer domain boundary in a dense phase monolayer 
could be modulated into a different form in less dense phases.}

\setcounter{equation}{0}
\setcounter{figure}{0}

\renewcommand{\theequation}{A\arabic{equation}}
\renewcommand{\thefigure}{A\arabic{figure}} 

\section{Appendix}

\begin{figure}
\includegraphics[width=1.0\linewidth]{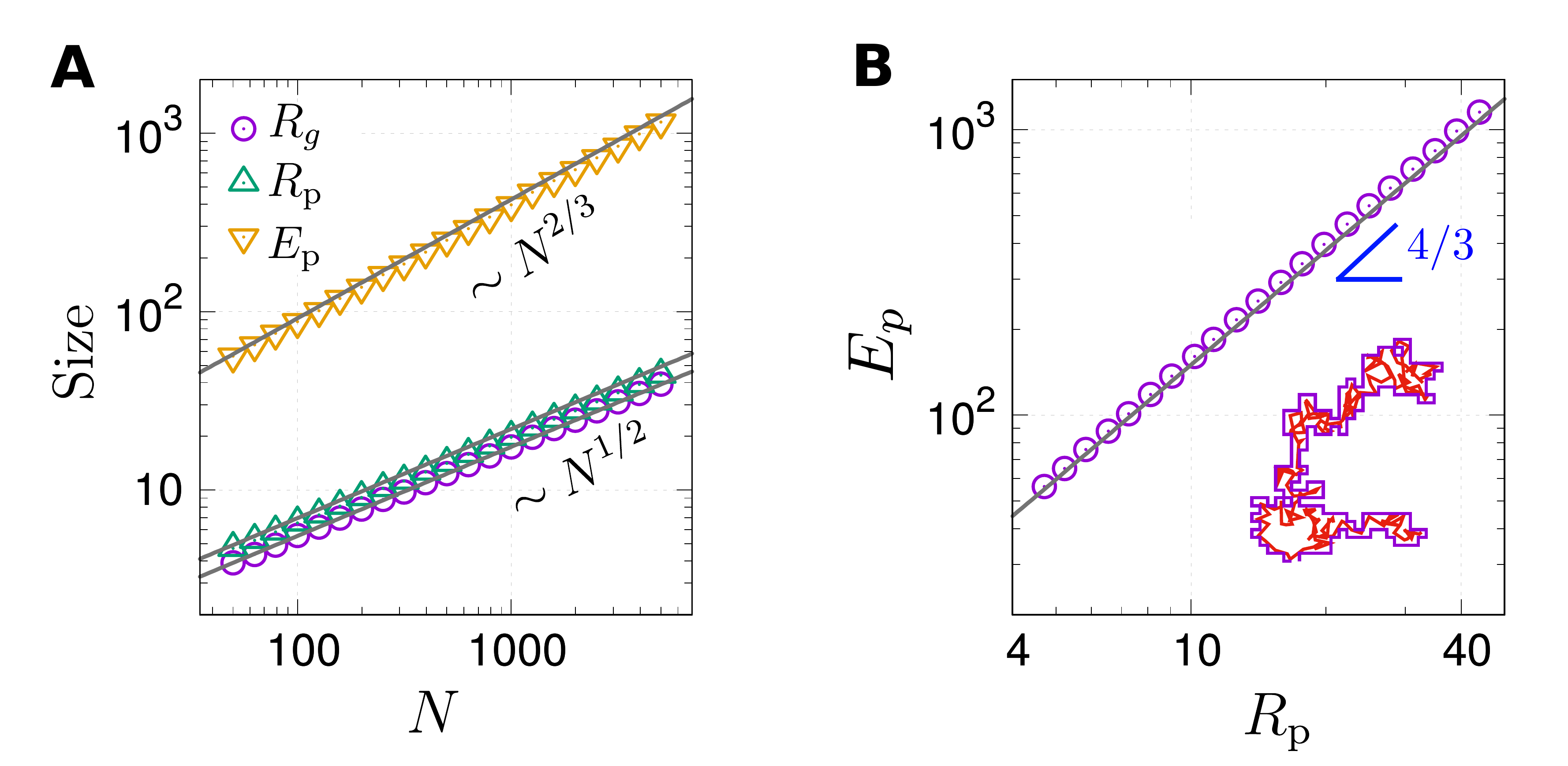}
\caption{
Hausdorff dimension of planar Brownian frontier.
(A) Scaling relations of $R_p$ (or $R_g$) with $N$, and $E_p$ with $N$ obtained from simulations of 2D Brownian random walks with varying $N$. 
(B) $E_p$ versus $R_p$, which determines the scaling exponent of the external perimeter of 2D Brownian random walks, $\mathcal{D}^\partial_B=4/3$.}
\label{planarBrown}
\end{figure}

\subsection{Hausdorff dimension of planar Brownian frontier}  
For the sanity check of the fractal dimension $\mathcal{D}^\partial$ obtained from our study, we generated the Brownian motion in 2D plane. 
First, $R_p(N)$, the radius of gyration of the monomers constituting the external perimeter of Brownian motion (Brownian frontier), displays effectively an identical scaling relation with the radius of gyration $R_g(N)$ of the full chain as $R_p\sim R_g\sim N^{1/2}$.  
Further, we examine the variation of the external perimeter as a function of $N$, finding that $E_p$ satisfies the scaling of $E_p\sim N^{2/3}$ (Fig.~\ref{planarBrown}A). 
Therefore, from the relation of $E_p\sim R_p^{\mathcal{D}^\partial}$ (Eq.~\ref{eqn:fractal}), we obtain $\mathcal{D}^\partial=4/3$. 
This indicates that the fractal (Haudorff) dimension of the external perimeter of planar Brownian motion ($\mathcal{D}^\partial_{\rm B}$) is identical to the fractal dimension of 2D SAW ($\mathcal{D}_{\rm SAW}=(\nu_{\rm SAW}^{2D})^{-1}$), i.e., $\mathcal{D}^\partial_{\rm B}=\mathcal{D}_{\rm SAW}=4/3$.

\subsection{Polymer-magnet analogy and critical exponents}  
Configurations of a linear polymer made of $N$ links are characterized by two critical exponents, $\nu$ and $\gamma$ \cite{deGennesbook}. 
The exponent $\nu$ is associated with the size (Flory radius, $R_F$) of a polymer chain 
that scales with $N$ as $R_F\sim N^{\nu}$, and $\gamma$ is asymptotically associated with the partition sum (the total number of configurations) $Z_N({\rm tot})\sim \mu^NN^{\gamma-1}$, where $\mu$ is the connectivity  constant that depends on the detail of the model.
Specifically, $\mu=6$ with $\gamma=1$ for the ideal chain in a cubic lattice in 3D, and $\mu=\sqrt{2+\sqrt{2}}$ with $\gamma=43/32$ for self-avoiding walk on a honeycomb lattice in 2D \cite{nienhuis1982exact,duplantier1986PRL,duminil2012connective} (see also Eq.\ref{xL_SAW}).

According to the polymer-ferromagnet analogy, which maps the chain statistics of polymer onto the $O(n)$ model of magnetic spin system  
defined by the energy hamiltonian 
\begin{align}
\mathcal{H}/\tau=-(K/\tau)\sum_{\langle i,j\rangle}\vec{S}_i\cdot \vec{S}_j
\label{eqn:hamiltonian}
\end{align} 
with the coupling constant between the neighboring spins $K>0$ and $\sum_{\alpha=1}^nS_{i\alpha}^2=n$~\cite{deGennes72PhysLett,deGennesbook},   
the correlation between $i$-th and $j$-th spins, near the critical temperature $\tau=\tau_c(1+\epsilon)\simeq \tau_ce^{\epsilon}$ and at the limit of $n=0$, is equivalent to the number of self-avoiding configurations $Z_N(ij)$ weighted by a factor $(K/\tau)^N$, i.e., $\langle S_{i\alpha}S_{j\alpha}\rangle\Big|_{n=0}=\sum_NZ_N(ij)(K/\tau)^N$~\cite{deGennes72PhysLett,deGennesbook}. 
Along with the definition of partition sum $Z_N({\rm tot})=\sum_jZ_N(ij)$, 
the magnetic susceptibility (or the two-point correlation) can be associated with $Z_N({\rm tot})$ as follows: 
\begin{align}
\chi_M&\simeq \frac{1}{\tau}\sum_j\langle S_{i\alpha}S_{j\alpha}\rangle\Big|_{n=0}=\frac{1}{\tau}\sum_NZ_N({\rm tot}) (K/\tau)^N\nonumber\\
&\sim \tau^{-1}\int_0^{\infty}dNe^{-\epsilon N}N^{\gamma-1}\sim |\epsilon|^{-\gamma}, 
\end{align}
where $\mu K=\tau_c$ is used in the second line.  
This clarifies that the exponent $\gamma$ characterizing $Z_N({\rm tot})$ is the same exponent associated with the  susceptibility near the critical temperature. 
It also shows that $N$ is the conjugate variable of $\epsilon$ ($N\sim \epsilon^{-1}$), which enables to map the polymer size ($R_F\sim N^\nu$) to the correlation length of the magnetic system ($\xi\sim |\epsilon|^{-\nu}$) as long as $N$ is sufficiently large. 
{\color{black}In other words, a polymer chain with $N\gg 1$ can be considered as a scale-invariant, critical object.}

\begin{figure}[t]
\includegraphics[width=0.8\linewidth]{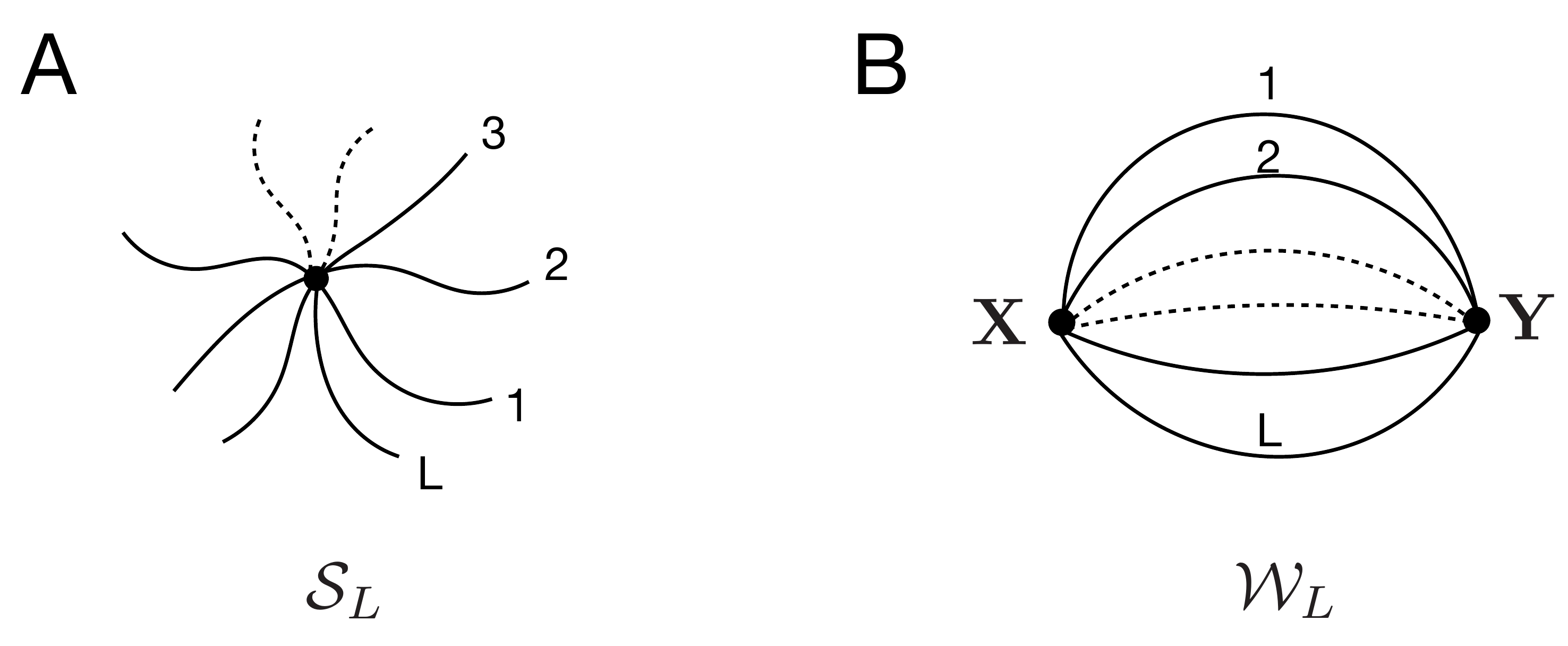}
\caption{The graphs of (A) a $L$-arm star ($\mathcal{G}=\mathcal{S}_L$) and (B) a watermelon geometry ($\mathcal{G}=\mathcal{W}_L$).  
}
\label{graph}
\end{figure}

The foregoing argument for the linear polymer can be extended to a polymer system with more complicated geometry represented using a general graph $\mathcal{G}$.  
For a general graph $\mathcal{G}$ made of $\mathcal{N}$ chains of an identical length $N$ comprising $n_L$ $L$-arm vertices, 
the partition sum $\mathcal{Z}(\mathcal{G})$ scales with $N$ as $\mathcal{Z}(\mathcal{G})\sim N^{\gamma_{\mathcal{G}}-1}$, and the enhancement exponent $\gamma_{\mathcal{G}}$ satisfies the following hyperscaling relation \cite{duplantier1986PRL,Duplantier89JSP} 
\begin{align}
\gamma_\mathcal{G}-1=\sum_{L\geq 1}n_L\sigma_L-\nu d\mathcal{L}, 
\label{eqn:hyper}
\end{align}
where 
$\mathcal{N}$ is related to $n_L$ as $\mathcal{N}=\frac{1}{2}\sum_{L\geq 1}n_LL$, 
$\mathcal{L}$ denotes the number of loops in the graph $\mathcal{G}$, $\mathcal{L}=\sum_{L\geq 1}(L-2)n_L/2+1$, 
and $\sigma_L$ is the exponent associated with the $L$-vertex.
\begin{itemize}
\item For $L$-arm star polymers ($n_1=L$, $n_L=1$, and $n_{L\neq1,L}=0$) ($\mathcal{G}=\mathcal{S}_L$, Fig.~\ref{graph}-A), 
one gets $\mathcal{L}=0$ and 
\begin{align}
\gamma_{\mathcal{S}_L}-1=L\sigma_1+\sigma_L.
\label{eqn:hyper_L}
\end{align}
\item For a graph of the watermelon geometry ($n_L=2$,  $n_{L'\neq L}=0$, and $\mathcal{L}=L-1$) ($\mathcal{G}=\mathcal{W}_L$, Fig.~\ref{graph}B) with $L$ arms of the same length, Eq.\ref{eqn:hyper} yields 
\begin{align}
\gamma_{\mathcal{W}_L}-1=2\sigma_L-\nu d(L-1). 
\label{eqn:watermelon_gamma}
\end{align}
\end{itemize}

{\color{black}In fact, the enhancement exponent of $L$-arm star polymer $\gamma_L(\equiv \gamma_{\mathcal{S}_L})$ can be related with the scaling dimension $x_L$ via $\sigma_L$, where 
$x_L$ is the exponent describing the algebraic decay of two-point correlation between $L$-arm vertices at ${\bf X}$ and ${\bf Y}$ at the critical point $\beta_c$: 
\begin{align}
\langle\varphi_L({\bf X})\varphi_L({\bf Y})\rangle_{\beta_c}=\frac{1}{|{\bf X}-{\bf Y}|^{2x_L}}. 
\end{align} 
To derive the relation between $\sigma_L$ and $x_L$, 
one considers the two-point correlation $\langle\varphi_L({\bf X})\varphi_L({\bf Y})\rangle_\beta$ at a temperature $\beta^{-1}$ and expresses it in terms of the partition sum of all the possible watermelon-like configurations between ${\bf X}$ and ${\bf Y}$ \cite{duplantier1989PR,Duplantier89JSP,duplantier1986PRL},}  
\begin{widetext}
\begin{align}
\langle\varphi_L({\bf X})\varphi_L({\bf Y})\rangle_\beta=\int_0^{\infty}\prod_{\alpha=1}^LdN_\alpha e^{-\beta(N_1+N_2+\cdots+N_L)}\mathcal{Z}_L(N_1,\ldots,N_L,{\bf X},{\bf Y}). 
\end{align}
Then, its inverse Laplace transform is written as 
\begin{align}
\frac{1}{2\pi i}\int_{\sigma-i\infty}^{\sigma+i\infty}d\beta e^{\beta LN}\langle\varphi_L({\bf X})\varphi_L({\bf Y})\rangle_\beta=\int_0^\infty\prod_{\alpha=1}^L dN_\alpha\delta\left(LN-\sum_{\alpha=1}^LN_\alpha\right)\mathcal{Z}_L(N_1,\ldots,N_L,{\bf X},{\bf Y}).
\label{corr} 
\end{align}
Integrating the right hand side of Eq.\ref{corr} over the space $\int d{\bf Y}(\ldots)$ defines the partition sum of polydisperse $L$-arm star polymers $\mathcal{Z}_L^{\rm poly}(N)$. 
Thus, at the critical point $\beta=\beta_c$, one obtains   
\begin{align}
\mathcal{Z}^{\rm poly}_L(N)=\frac{1}{2\pi i}\int d\beta e^{\beta LN}\int d^d{\bf Y}\langle\varphi_L({\bf X})\varphi_L({\bf Y})\rangle_{\beta_c}\sim
(LN)^{-1}\underbrace{\int d^d{\bf Y}\frac{1}{|{\bf X}-{\bf Y}|^{2x_L}}}_{\sim R^{d-2x_L}\sim N^{\nu(d-2x_L)}},  
\label{eqn:poly}
\end{align}
\end{widetext}
where we have used the {\color{black}Flory's relation} $R\sim N^{\nu}$, and obtain 
\begin{align}
\mathcal{Z}_L^{\rm poly}(N)\sim N^{\nu(d-2x_L)-1}. 
\label{Zpoly}
\end{align} 
Meanwhile, the partition sum of the watermelon type graphs with $L$ arms of the same length $N_\alpha=N$, $\mathcal{Z}(\mathcal{W}_L)\equiv \mathcal{Z}_L(N,N,\cdots ,N)$, 
is related with $\mathcal{Z}_L^{\rm poly}(N)$ as~\cite{duplantier1986PRL}
\begin{align}
\mathcal{Z}_L^{\rm poly}(N)&\sim \mathcal{Z}(\mathcal{W}_L)N^{L-1}\nonumber\\
&\sim N^{\gamma_{\mathcal{W}_L}-1}N^{L-1}. 
\label{Zpoly2}
\end{align}
Therefore, the relation between $\sigma_L$ and $x_L$ follows from 
Eqs.~\ref{Zpoly} and \ref{Zpoly2} along with Eq.\ref{eqn:watermelon_gamma}   
\begin{align}
\sigma_L=-\nu x_L+L(\nu d-1)/2. 
\label{eqn:vertex_scalingdim}
\end{align}

\subsection{Contact exponent}
The interchain contact probability $f_{\rm inter}$ discussed in Eq.~\ref{eq:Ep} is calculated by means of the contact exponent $\theta\equiv \theta_{2,2}$ of the distribution function at small separation $r\simeq a$,   
$P(r)\sim \frac{1}{R_F^{d\mathcal{L}}}\left(\frac{r}{R_F}\right)^{\theta}$ \cite{duplantier1987PRB,Duplantier89JSP} with $\mathcal{L}=0$, that is, 
\begin{align}
f_{\rm inter}&\sim\left(\frac{a}{R_F}\right)^{\theta}\sim N^{-\nu\theta}
\label{eqn:f_inter}
\end{align}
Therefore, comparison of Eq.\ref{eqn:f_inter} with $f_{\rm inter}=N^{\gamma_4-2\gamma_2+1}$ yields 
\begin{align}
\theta=\frac{1}{\nu}(2\gamma_2-\gamma_4-1) 
\end{align}
and $\mathcal{D}^{\partial}_{\rm D}=1/\nu-\theta$ from $E_p\sim N^{\nu\mathcal{D}^{\partial}_{\rm D}}\sim N^{1-\nu\theta}$, giving rise to Eq.\ref{DD}.  

More generally, the contact exponent between $L$-arm and $L'$-arm star polymers, $\theta_{L,L'}$, can be related with other critical exponents for $L$- and $L'$-arm star polymers using Eq.~\ref{eqn:hyper_L} and Eq.~\ref{eqn:vertex_scalingdim} \cite{Duplantier89JSP}
\begin{align}
\theta_{L,L'}&=\frac{1}{\nu}(\gamma_L+\gamma_{L'}-\gamma_{L+L'}-1)\nonumber\\
&=\frac{1}{\nu}(\sigma_L+\sigma_{L'}-\sigma_{L+L'})\nonumber\\
&=x_{L+L'}-x_L-x_{L'}. 
\label{eqn:theta}
\end{align}

For polymers in a dense phase, 
$x_L=(L^2-4)/16$ (see Eq.~\ref{xL_d}), which leads to 
$\theta_{2,2}=\theta=x_4-2x_2=3/4$, and hence 
$\mathcal{D}^{\partial}_{\rm D}=1/\nu-\theta=5/4$~\cite{meyer2009perimeter}.

\subsection{The universal scaling dimensions from mapping spin models onto the 2D Coulomb gas model} 
For the aforementioned $O(n)$ model (Eq.\ref{eqn:hamiltonian}), 
the corresponding partition function\\  
$Z_{O(n)}=\int \prod_kdS_k\prod_{\langle i,j\rangle}(1+tS_i\cdot S_j)$   can be expressed as a sum of the diagrams of closed loops~\cite{nienhuis1982exact},  
\begin{align}
Z_{O(n)}=\sum_{\mathcal{G}}t^{\mathcal{N}_B(\mathcal{G})}n^{\mathcal{N}_C(\mathcal{G})}
\end{align}
where $t\equiv e^{K/\tau}-1$, and $\mathcal{N}_B(\mathcal{G})$ and $\mathcal{N}_C(\mathcal{G})$ are the total number of bonds and non-intersecting loops of the graph $\mathcal{G}$. 
Considering closed loop configurationss in a triangular solid-on-solid (TSOS) model defined on a honeycomb lattice, one can identify the condition that associates $O(n)$ model with the TSOS model. 
Since the partition function of the TSOS model is given as $Z_{\rm TSOS}=\sum_\mathcal{G}t^{\mathcal{N}_B(\mathcal{G})}(e^{6i\alpha}+e^{-6i\alpha})^{\mathcal{N}_C(\mathcal{G})}$ where $(e^{6i\alpha}+e^{-6i\alpha})$ is the factor contributed by a closed loop on the hexagonal lattice, 
Thus, the invariance of partition function of the two alternative models ($Z_{O(n)}=Z_{\rm TSOS}$) yields the relation 
$n=2\cos{6\alpha}$. 
On a hexagonal lattice, $t_c=(2+(2-n)^{1/2})^{-1/2}<1$ is the exact critical point of the $O(n)$ model, and $t=1$ corresponds to the critical low-temperature phase $t>t_c$~\cite{nienhuis1982exact}. 
Next, mapping the TSOS model onto the 2D Coulomb gas model via $6\alpha/\pi=(1-g)$~\cite{nienhuis1984critical}  
allows one to relate $n$ of the $O(n)$-model with the coupling strength of Coulomb gas ($g$), whose action in the continuum limit is $\mathcal{A}=\frac{g}{4\pi}\int(\nabla \varphi)^2d^2{\bf r}$~\cite{duplantier1989PR}, as follows. 
\begin{align}
n=-2\cos{(\pi g)}.
\label{eqn:n_to_g}
\end{align}
For the case of SAW chain ($n=0$), $g=3/2\in[1,2]$ at dilute phases ($t=t_c$), whereas $g=1/2\in[0,1]$ at dense phases ($t>t_c$)~\cite{duplantier1987exact,duplantier1989PR,duplantier1986JPA}. 
For the $\Theta$ chain, whose statistics is identical to the percolation ($n=1$), Eq.~\ref{eqn:n_to_g} leads to $g=2/3$ \cite{Duplantier87PRL}.

For the 2D Columb gas model, the two-point correlation between ${\bf X}$ and ${\bf Y}$ at the critical point, $\langle\varphi_L({\bf X})\varphi_L({\bf Y})\rangle_{\beta_c}=|{\bf X}-{\bf Y}|^{-2x_L}$, is contributed by vortex configurations and spin waves~\cite{kosterlitz1973ordering,kosterlitz1974critical} and it can be interpreted as an outcome of the correlation due to magnetic and electric charges~\cite{jose1977renormalization,nienhuis1984critical,duplantier1987critical,duplantier1989PR} 
\begin{align}
\langle\varphi_L({\bf X})\varphi_L({\bf Y})\rangle_{\beta_c}=|{\bf X}-{\bf Y}|^{gm_{\bf X}m_{\bf Y}+e_{\bf X}e_{\bf Y}/g}. 
\label{eqn:LL}
\end{align}
For the correlation between two $L$-arm vertices of the watermelon geometry, the magnetic charges at ${\bf X}$ and ${\bf Y}$ are due to a vortex and anti-vortex pair with $L/2$ dislocation{\color{black}s} $m_{\bf X}=-m_{\bf Y}=L/2$,  
and the electric charges at ${\bf X}$ and ${\bf Y}$ are given as $e_{\bf X}=e_{\bf Y}=1-g$, and hence the scaling dimension $x_L=-(gm_{\bf X}m_{\bf Y}+e_{\bf X}e_{\bf Y}/g)/2$ is expressed in terms of $g$ and $L$~\cite{duplantier1987critical}
\begin{align}
x_L=\frac{L^2}{8}g-\frac{(g-1)^2}{2g}. 
\end{align}

Taken together, along with the $L$-arm vertex exponent ($\sigma_L=-\nu x_L+L(\nu d-1)/2$ for $L\geq 1$), 
the scaling dimensions ($x_L$) for $L$-arm star polymers under three different conditions are obtained as follows~\cite{duplantier1989PR,Duplantier89JSP}. 
\begin{itemize}
\item SAW chain in dilute phase ($n=0$, $g=3/2$): 
\begin{align}
x^{\text{SAW}}_L=\frac{(9L^2-4)}{48}.
\label{xL_SAW}
\end{align}
Thus, from Eq.\ref{eqn:vertex_scalingdim} with $\nu=3/4$ and $d=2$, one gets the $L$-arm vertex exponent~\cite{duplantier1986PRL}
\begin{align}
\sigma_L^{\rm SAW}=\frac{(2-L)(9L+2)}{64},  
\end{align}
and from Eq.\ref{eqn:hyper_L}
\begin{align}
\gamma_L^{\rm SAW}=\frac{[68+9L(3-L)]}{64}. 
\label{eqn:gamma_SAW}
\end{align}
Thus, Eq.\ref{eqn:gamma_SAW} confirms the Nienhuis result $\gamma^{\rm SAW}\equiv\gamma_1=\gamma_2=43/32$ \cite{nienhuis1982exact}.

\item Polymer solution in dense phase ($n=0$, $g=1/2$): 
\begin{align}
x^{\rm D}_L=\frac{(L^2-4)}{16}, 
\label{xL_d}
\end{align}
and with $\nu=1/2$ and $d=2$ 
\begin{align}
\sigma^{\rm D}_L=-\frac{(L^2-4)}{32}. 
\end{align}
Since $\theta_{2,2}=x_4^{\rm D}-2x_2^{\rm D}$ with $x_4^{\rm D}=3/4$ and $x_2^{\rm D}=0$, one gets $\theta_{2,2}=3/4$.   
Thus, the relation  $\gamma_L^{\rm D}=1+\sigma_L^{\rm D}+L\sigma_1^{\rm D}$ (Eq.~\ref{eqn:hyper_L}) offers the enhancement exponent for $L$-arm polymer in dense phase 
\begin{align}
\gamma_L^{\rm D}=\frac{9}{8}+\frac{(3-L)L}{32},  
\label{eqn:enhancement_D}
\end{align}
which confirms $\gamma^{\rm D}=19/16$ for linear polymers~\cite{duplantier1986JPA}. 
\item $\Theta$ chain in dilute phase ($n=1$, $g=2/3$): 
\begin{align}
x_L^{\Theta}=\frac{(L^2-1)}{12}, 
\label{xL_Theta}
\end{align}
and with $\nu=4/7$ and $d=2$  
\begin{align}
\sigma^{\Theta}_L=-\frac{(2L+1)(L-2)}{42}. 
\end{align}

\end{itemize}

\acknowledgements{
This work was in part supported by National Natural Science Foundation of China, ZSTU intramural grant (12104404, 20062226-Y to L.L.), and KIAS Individual Grant (CG035003 to C.H.) at Korea Institute for Advanced Study. 
C.H. is grateful to Dr. Daeseong Yong for careful reading of the manuscript and helpful comments. 
We thank the Center for Advanced Computation in KIAS for providing computing resources. 
}

\bibliography{mybib1}

\begin{thebibliography}{66}%
\makeatletter
\providecommand \@ifxundefined [1]{%
 \@ifx{#1\undefined}
}%
\providecommand \@ifnum [1]{%
 \ifnum #1\expandafter \@firstoftwo
 \else \expandafter \@secondoftwo
 \fi
}%
\providecommand \@ifx [1]{%
 \ifx #1\expandafter \@firstoftwo
 \else \expandafter \@secondoftwo
 \fi
}%
\providecommand \natexlab [1]{#1}%
\providecommand \enquote  [1]{``#1''}%
\providecommand \bibnamefont  [1]{#1}%
\providecommand \bibfnamefont [1]{#1}%
\providecommand \citenamefont [1]{#1}%
\providecommand \href@noop [0]{\@secondoftwo}%
\providecommand \href [0]{\begingroup \@sanitize@url \@href}%
\providecommand \@href[1]{\@@startlink{#1}\@@href}%
\providecommand \@@href[1]{\endgroup#1\@@endlink}%
\providecommand \@sanitize@url [0]{\catcode `\\12\catcode `\$12\catcode
  `\&12\catcode `\#12\catcode `\^12\catcode `\_12\catcode `\%12\relax}%
\providecommand \@@startlink[1]{}%
\providecommand \@@endlink[0]{}%
\providecommand \url  [0]{\begingroup\@sanitize@url \@url }%
\providecommand \@url [1]{\endgroup\@href {#1}{\urlprefix }}%
\providecommand \urlprefix  [0]{URL }%
\providecommand \Eprint [0]{\href }%
\providecommand \doibase [0]{https://doi.org/}%
\providecommand \selectlanguage [0]{\@gobble}%
\providecommand \bibinfo  [0]{\@secondoftwo}%
\providecommand \bibfield  [0]{\@secondoftwo}%
\providecommand \translation [1]{[#1]}%
\providecommand \BibitemOpen [0]{}%
\providecommand \bibitemStop [0]{}%
\providecommand \bibitemNoStop [0]{.\EOS\space}%
\providecommand \EOS [0]{\spacefactor3000\relax}%
\providecommand \BibitemShut  [1]{\csname bibitem#1\endcsname}%
\let\auto@bib@innerbib\@empty
\bibitem [{\citenamefont {{de Gennes}}(1979)}]{deGennesbook}%
  \BibitemOpen
  \bibfield  {author} {\bibinfo {author} {\bibfnamefont {P.~G.}\ \bibnamefont
  {{de Gennes}}},\ }\href@noop {} {\emph {\bibinfo {title} {{Scaling Concepts
  in Polymer Physics}}}}\ (\bibinfo  {publisher} {Cornell University Press},\
  \bibinfo {address} {Ithaca and London},\ \bibinfo {year} {1979})\BibitemShut
  {NoStop}%
\bibitem [{\citenamefont {Meirovitch}\ \emph {et~al.}(1989)\citenamefont
  {Meirovitch}, \citenamefont {Chang},\ and\ \citenamefont
  {Shapir}}]{meirovitch1989surface}%
  \BibitemOpen
  \bibfield  {author} {\bibinfo {author} {\bibfnamefont {H.}~\bibnamefont
  {Meirovitch}}, \bibinfo {author} {\bibfnamefont {I.}~\bibnamefont {Chang}},\
  and\ \bibinfo {author} {\bibfnamefont {Y.}~\bibnamefont {Shapir}},\
  }\bibfield  {title} {\bibinfo {title} {Surface exponents of trails in two
  dimensions at tricriticality: Computer simulation study},\ }\href@noop {}
  {\bibfield  {journal} {\bibinfo  {journal} {Phys. Rev. A}\ }\textbf {\bibinfo
  {volume} {40}},\ \bibinfo {pages} {2879} (\bibinfo {year}
  {1989})}\BibitemShut {NoStop}%
\bibitem [{\citenamefont {Semenov}\ and\ \citenamefont
  {Johner}(2003)}]{semenov2003EPJE}%
  \BibitemOpen
  \bibfield  {author} {\bibinfo {author} {\bibfnamefont {A.}~\bibnamefont
  {Semenov}}\ and\ \bibinfo {author} {\bibfnamefont {A.}~\bibnamefont
  {Johner}},\ }\bibfield  {title} {\bibinfo {title} {Theoretical notes on dense
  polymers in two dimensions},\ }\href@noop {} {\bibfield  {journal} {\bibinfo
  {journal} {Eur. Phys. J. E}\ }\textbf {\bibinfo {volume} {12}},\ \bibinfo
  {pages} {469} (\bibinfo {year} {2003})}\BibitemShut {NoStop}%
\bibitem [{\citenamefont {Luengo}\ \emph {et~al.}(1997)\citenamefont {Luengo},
  \citenamefont {Schmitt}, \citenamefont {Hill},\ and\ \citenamefont
  {Israelachvili}}]{luengo1997thin}%
  \BibitemOpen
  \bibfield  {author} {\bibinfo {author} {\bibfnamefont {G.}~\bibnamefont
  {Luengo}}, \bibinfo {author} {\bibfnamefont {F.-J.}\ \bibnamefont {Schmitt}},
  \bibinfo {author} {\bibfnamefont {R.}~\bibnamefont {Hill}},\ and\ \bibinfo
  {author} {\bibfnamefont {J.}~\bibnamefont {Israelachvili}},\ }\bibfield
  {title} {\bibinfo {title} {Thin film rheology and tribology of confined
  polymer melts: contrasts with bulk properties},\ }\href@noop {} {\bibfield
  {journal} {\bibinfo  {journal} {Macromolecules}\ }\textbf {\bibinfo {volume}
  {30}},\ \bibinfo {pages} {2482} (\bibinfo {year} {1997})}\BibitemShut
  {NoStop}%
\bibitem [{\citenamefont {Jones}(1999)}]{jones1999dynamics}%
  \BibitemOpen
  \bibfield  {author} {\bibinfo {author} {\bibfnamefont {R.}~\bibnamefont
  {Jones}},\ }\bibfield  {title} {\bibinfo {title} {The dynamics of thin
  polymer films},\ }\href@noop {} {\bibfield  {journal} {\bibinfo  {journal}
  {Current opinion in colloid \& interface science}\ }\textbf {\bibinfo
  {volume} {4}},\ \bibinfo {pages} {153} (\bibinfo {year} {1999})}\BibitemShut
  {NoStop}%
\bibitem [{\citenamefont {Katsumata}\ \emph {et~al.}(2018)\citenamefont
  {Katsumata}, \citenamefont {Dulaney}, \citenamefont {Kim},\ and\
  \citenamefont {Ellison}}]{katsumata2018glass}%
  \BibitemOpen
  \bibfield  {author} {\bibinfo {author} {\bibfnamefont {R.}~\bibnamefont
  {Katsumata}}, \bibinfo {author} {\bibfnamefont {A.~R.}\ \bibnamefont
  {Dulaney}}, \bibinfo {author} {\bibfnamefont {C.~B.}\ \bibnamefont {Kim}},\
  and\ \bibinfo {author} {\bibfnamefont {C.~J.}\ \bibnamefont {Ellison}},\
  }\bibfield  {title} {\bibinfo {title} {Glass transition and self-diffusion of
  unentangled polymer melts nanoconfined by different interfaces},\ }\href@noop
  {} {\bibfield  {journal} {\bibinfo  {journal} {Macromolecules}\ }\textbf
  {\bibinfo {volume} {51}},\ \bibinfo {pages} {7509} (\bibinfo {year}
  {2018})}\BibitemShut {NoStop}%
\bibitem [{\citenamefont {Frank}\ \emph {et~al.}(1996)\citenamefont {Frank},
  \citenamefont {Gast}, \citenamefont {Russell}, \citenamefont {Brown},\ and\
  \citenamefont {Hawker}}]{frank1996polymer}%
  \BibitemOpen
  \bibfield  {author} {\bibinfo {author} {\bibfnamefont {B.}~\bibnamefont
  {Frank}}, \bibinfo {author} {\bibfnamefont {A.~P.}\ \bibnamefont {Gast}},
  \bibinfo {author} {\bibfnamefont {T.~P.}\ \bibnamefont {Russell}}, \bibinfo
  {author} {\bibfnamefont {H.~R.}\ \bibnamefont {Brown}},\ and\ \bibinfo
  {author} {\bibfnamefont {C.}~\bibnamefont {Hawker}},\ }\bibfield  {title}
  {\bibinfo {title} {Polymer mobility in thin films},\ }\href@noop {}
  {\bibfield  {journal} {\bibinfo  {journal} {Macromolecules}\ }\textbf
  {\bibinfo {volume} {29}},\ \bibinfo {pages} {6531} (\bibinfo {year}
  {1996})}\BibitemShut {NoStop}%
\bibitem [{\citenamefont {Bay}\ \emph {et~al.}(2018)\citenamefont {Bay},
  \citenamefont {Shimomura}, \citenamefont {Liu}, \citenamefont {Ilton},\ and\
  \citenamefont {Crosby}}]{bay2018confinement}%
  \BibitemOpen
  \bibfield  {author} {\bibinfo {author} {\bibfnamefont {R.~K.}\ \bibnamefont
  {Bay}}, \bibinfo {author} {\bibfnamefont {S.}~\bibnamefont {Shimomura}},
  \bibinfo {author} {\bibfnamefont {Y.}~\bibnamefont {Liu}}, \bibinfo {author}
  {\bibfnamefont {M.}~\bibnamefont {Ilton}},\ and\ \bibinfo {author}
  {\bibfnamefont {A.~J.}\ \bibnamefont {Crosby}},\ }\bibfield  {title}
  {\bibinfo {title} {Confinement effect on strain localizations in glassy
  polymer films},\ }\href@noop {} {\bibfield  {journal} {\bibinfo  {journal}
  {Macromolecules}\ }\textbf {\bibinfo {volume} {51}},\ \bibinfo {pages} {3647}
  (\bibinfo {year} {2018})}\BibitemShut {NoStop}%
\bibitem [{\citenamefont {Liu}\ \emph {et~al.}(2019)\citenamefont {Liu},
  \citenamefont {Pincus},\ and\ \citenamefont {Hyeon}}]{Liu19Nanolett}%
  \BibitemOpen
  \bibfield  {author} {\bibinfo {author} {\bibfnamefont {L.}~\bibnamefont
  {Liu}}, \bibinfo {author} {\bibfnamefont {P.~A.}\ \bibnamefont {Pincus}},\
  and\ \bibinfo {author} {\bibfnamefont {C.}~\bibnamefont {Hyeon}},\ }\bibfield
   {title} {\bibinfo {title} {{Compressing $\Theta$-chain in slit geometry}},\
  }\href@noop {} {\bibfield  {journal} {\bibinfo  {journal} {Nano Lett.}\
  }\textbf {\bibinfo {volume} {19}},\ \bibinfo {pages} {5667} (\bibinfo {year}
  {2019})}\BibitemShut {NoStop}%
\bibitem [{\citenamefont {Pressly}\ \emph {et~al.}(2019)\citenamefont
  {Pressly}, \citenamefont {Riggleman},\ and\ \citenamefont
  {Winey}}]{pressly2019increased}%
  \BibitemOpen
  \bibfield  {author} {\bibinfo {author} {\bibfnamefont {J.~F.}\ \bibnamefont
  {Pressly}}, \bibinfo {author} {\bibfnamefont {R.~A.}\ \bibnamefont
  {Riggleman}},\ and\ \bibinfo {author} {\bibfnamefont {K.~I.}\ \bibnamefont
  {Winey}},\ }\bibfield  {title} {\bibinfo {title} {Increased polymer
  diffusivity in thin-film confinement},\ }\href@noop {} {\bibfield  {journal}
  {\bibinfo  {journal} {Macromolecules}\ }\textbf {\bibinfo {volume} {52}},\
  \bibinfo {pages} {6116} (\bibinfo {year} {2019})}\BibitemShut {NoStop}%
\bibitem [{\citenamefont {Langevin}\ and\ \citenamefont
  {Monroy}(2010)}]{langevin2010interfacial}%
  \BibitemOpen
  \bibfield  {author} {\bibinfo {author} {\bibfnamefont {D.}~\bibnamefont
  {Langevin}}\ and\ \bibinfo {author} {\bibfnamefont {F.}~\bibnamefont
  {Monroy}},\ }\bibfield  {title} {\bibinfo {title} {Interfacial rheology of
  polyelectrolytes and polymer monolayers at the air--water interface},\
  }\href@noop {} {\bibfield  {journal} {\bibinfo  {journal} {Curr. Opin.
  Colloid \& interface Sci.}\ }\textbf {\bibinfo {volume} {15}},\ \bibinfo
  {pages} {283} (\bibinfo {year} {2010})}\BibitemShut {NoStop}%
\bibitem [{\citenamefont {Duplantier}(1986{\natexlab{a}})}]{duplantier1986JPA}%
  \BibitemOpen
  \bibfield  {author} {\bibinfo {author} {\bibfnamefont {B.}~\bibnamefont
  {Duplantier}},\ }\bibfield  {title} {\bibinfo {title} {Exact critical
  exponents for two-dimensional dense polymers},\ }\href@noop {} {\bibfield
  {journal} {\bibinfo  {journal} {J. Phys. A: Math. Gen.}\ }\textbf {\bibinfo
  {volume} {19}},\ \bibinfo {pages} {L1009} (\bibinfo {year}
  {1986}{\natexlab{a}})}\BibitemShut {NoStop}%
\bibitem [{\citenamefont {Saleur}\ and\ \citenamefont
  {Duplantier}(1987)}]{saleur1987PRL}%
  \BibitemOpen
  \bibfield  {author} {\bibinfo {author} {\bibfnamefont {H.}~\bibnamefont
  {Saleur}}\ and\ \bibinfo {author} {\bibfnamefont {B.}~\bibnamefont
  {Duplantier}},\ }\bibfield  {title} {\bibinfo {title} {Exact determination of
  the percolation hull exponent in two dimensions},\ }\href@noop {} {\bibfield
  {journal} {\bibinfo  {journal} {Phys. Rev. Lett.}\ }\textbf {\bibinfo
  {volume} {58}},\ \bibinfo {pages} {2325} (\bibinfo {year}
  {1987})}\BibitemShut {NoStop}%
\bibitem [{\citenamefont
  {Duplantier}(1987{\natexlab{a}})}]{duplantier1987critical}%
  \BibitemOpen
  \bibfield  {author} {\bibinfo {author} {\bibfnamefont {B.}~\bibnamefont
  {Duplantier}},\ }\bibfield  {title} {\bibinfo {title} {{Critical exponents of
  Manhattan Hamiltonian walks in two dimensions, from Potts and $O(n)$
  models}},\ }\href@noop {} {\bibfield  {journal} {\bibinfo  {journal} {J.
  Stat. Phys.}\ }\textbf {\bibinfo {volume} {49}},\ \bibinfo {pages} {411}
  (\bibinfo {year} {1987}{\natexlab{a}})}\BibitemShut {NoStop}%
\bibitem [{\citenamefont {Duplantier}\ and\ \citenamefont
  {Saleur}(1987{\natexlab{a}})}]{duplantier1987exact}%
  \BibitemOpen
  \bibfield  {author} {\bibinfo {author} {\bibfnamefont {B.}~\bibnamefont
  {Duplantier}}\ and\ \bibinfo {author} {\bibfnamefont {H.}~\bibnamefont
  {Saleur}},\ }\bibfield  {title} {\bibinfo {title} {Exact critical properties
  of two-dimensional dense self-avoiding walks},\ }\href@noop {} {\bibfield
  {journal} {\bibinfo  {journal} {Nucl. Phys. B}\ }\textbf {\bibinfo {volume}
  {290}},\ \bibinfo {pages} {291} (\bibinfo {year}
  {1987}{\natexlab{a}})}\BibitemShut {NoStop}%
\bibitem [{\citenamefont {Duplantier}(1989{\natexlab{a}})}]{duplantier1989PR}%
  \BibitemOpen
  \bibfield  {author} {\bibinfo {author} {\bibfnamefont {B.}~\bibnamefont
  {Duplantier}},\ }\bibfield  {title} {\bibinfo {title} {Two-dimensional
  fractal geometry, critical phenomena and conformal invariance},\ }\href@noop
  {} {\bibfield  {journal} {\bibinfo  {journal} {Phys. Rep.}\ }\textbf
  {\bibinfo {volume} {184}},\ \bibinfo {pages} {229} (\bibinfo {year}
  {1989}{\natexlab{a}})}\BibitemShut {NoStop}%
\bibitem [{\citenamefont {Duplantier}(1987{\natexlab{b}})}]{duplantier1987PRB}%
  \BibitemOpen
  \bibfield  {author} {\bibinfo {author} {\bibfnamefont {B.}~\bibnamefont
  {Duplantier}},\ }\bibfield  {title} {\bibinfo {title} {Exact contact critical
  exponents of a self-avoiding polymer chain in two dimensions},\ }\href@noop
  {} {\bibfield  {journal} {\bibinfo  {journal} {Phys. Rev. B}\ }\textbf
  {\bibinfo {volume} {35}},\ \bibinfo {pages} {5290} (\bibinfo {year}
  {1987}{\natexlab{b}})}\BibitemShut {NoStop}%
\bibitem [{\citenamefont {Duplantier}(1989{\natexlab{b}})}]{Duplantier89JSP}%
  \BibitemOpen
  \bibfield  {author} {\bibinfo {author} {\bibfnamefont {B.}~\bibnamefont
  {Duplantier}},\ }\bibfield  {title} {\bibinfo {title} {Statistical mechanics
  of polymer networks of any topology},\ }\href@noop {} {\bibfield  {journal}
  {\bibinfo  {journal} {J. Stat. Phys.}\ }\textbf {\bibinfo {volume} {54}},\
  \bibinfo {pages} {581} (\bibinfo {year} {1989}{\natexlab{b}})}\BibitemShut
  {NoStop}%
\bibitem [{\citenamefont {Duplantier}\ and\ \citenamefont
  {Saleur}(1987{\natexlab{b}})}]{Duplantier87PRL}%
  \BibitemOpen
  \bibfield  {author} {\bibinfo {author} {\bibfnamefont {B.}~\bibnamefont
  {Duplantier}}\ and\ \bibinfo {author} {\bibfnamefont {H.}~\bibnamefont
  {Saleur}},\ }\bibfield  {title} {\bibinfo {title} {{Exact Tricritical
  Exponents for Polymers at the $\Theta$ Point in Two Dimensions}},\
  }\href@noop {} {\bibfield  {journal} {\bibinfo  {journal} {Phys. Rev. Lett.}\
  }\textbf {\bibinfo {volume} {59}},\ \bibinfo {pages} {539} (\bibinfo {year}
  {1987}{\natexlab{b}})}\BibitemShut {NoStop}%
\bibitem [{\citenamefont {Carmesin}\ and\ \citenamefont
  {Kremer}(1990)}]{carmesin1990JP}%
  \BibitemOpen
  \bibfield  {author} {\bibinfo {author} {\bibfnamefont {I.}~\bibnamefont
  {Carmesin}}\ and\ \bibinfo {author} {\bibfnamefont {K.}~\bibnamefont
  {Kremer}},\ }\bibfield  {title} {\bibinfo {title} {Static and dynamic
  properties of two-dimensional polymer melts},\ }\href@noop {} {\bibfield
  {journal} {\bibinfo  {journal} {J. Phys.}\ }\textbf {\bibinfo {volume}
  {51}},\ \bibinfo {pages} {915} (\bibinfo {year} {1990})}\BibitemShut
  {NoStop}%
\bibitem [{\citenamefont {Meyer}\ \emph {et~al.}(2009)\citenamefont {Meyer},
  \citenamefont {Kreer}, \citenamefont {Aichele}, \citenamefont {Cavallo},
  \citenamefont {Johner}, \citenamefont {Baschnagel},\ and\ \citenamefont
  {Wittmer}}]{meyer2009perimeter}%
  \BibitemOpen
  \bibfield  {author} {\bibinfo {author} {\bibfnamefont {H.}~\bibnamefont
  {Meyer}}, \bibinfo {author} {\bibfnamefont {T.}~\bibnamefont {Kreer}},
  \bibinfo {author} {\bibfnamefont {M.}~\bibnamefont {Aichele}}, \bibinfo
  {author} {\bibfnamefont {A.}~\bibnamefont {Cavallo}}, \bibinfo {author}
  {\bibfnamefont {A.}~\bibnamefont {Johner}}, \bibinfo {author} {\bibfnamefont
  {J.}~\bibnamefont {Baschnagel}},\ and\ \bibinfo {author} {\bibfnamefont
  {J.}~\bibnamefont {Wittmer}},\ }\bibfield  {title} {\bibinfo {title}
  {Perimeter length and form factor in two-dimensional polymer melts},\
  }\href@noop {} {\bibfield  {journal} {\bibinfo  {journal} {Phys. Rev. E}\
  }\textbf {\bibinfo {volume} {79}},\ \bibinfo {pages} {050802} (\bibinfo
  {year} {2009})}\BibitemShut {NoStop}%
\bibitem [{\citenamefont {Meyer}\ \emph {et~al.}(2010)\citenamefont {Meyer},
  \citenamefont {Wittmer}, \citenamefont {Kreer}, \citenamefont {Johner},\ and\
  \citenamefont {Baschnagel}}]{meyer2010static}%
  \BibitemOpen
  \bibfield  {author} {\bibinfo {author} {\bibfnamefont {H.}~\bibnamefont
  {Meyer}}, \bibinfo {author} {\bibfnamefont {J.}~\bibnamefont {Wittmer}},
  \bibinfo {author} {\bibfnamefont {T.}~\bibnamefont {Kreer}}, \bibinfo
  {author} {\bibfnamefont {A.}~\bibnamefont {Johner}},\ and\ \bibinfo {author}
  {\bibfnamefont {J.}~\bibnamefont {Baschnagel}},\ }\bibfield  {title}
  {\bibinfo {title} {Static properties of polymer melts in two dimensions},\
  }\href@noop {} {\bibfield  {journal} {\bibinfo  {journal} {J. Chem. Phys.}\
  }\textbf {\bibinfo {volume} {132}},\ \bibinfo {pages} {184904} (\bibinfo
  {year} {2010})}\BibitemShut {NoStop}%
\bibitem [{\citenamefont {Schulmann}\ \emph {et~al.}(2013)\citenamefont
  {Schulmann}, \citenamefont {Meyer}, \citenamefont {Kreer}, \citenamefont
  {Cavallo}, \citenamefont {Johner}, \citenamefont {Baschnagel},\ and\
  \citenamefont {Wittmer}}]{schulmann2013PSSC}%
  \BibitemOpen
  \bibfield  {author} {\bibinfo {author} {\bibfnamefont {N.}~\bibnamefont
  {Schulmann}}, \bibinfo {author} {\bibfnamefont {H.}~\bibnamefont {Meyer}},
  \bibinfo {author} {\bibfnamefont {T.}~\bibnamefont {Kreer}}, \bibinfo
  {author} {\bibfnamefont {A.}~\bibnamefont {Cavallo}}, \bibinfo {author}
  {\bibfnamefont {A.}~\bibnamefont {Johner}}, \bibinfo {author} {\bibfnamefont
  {J.}~\bibnamefont {Baschnagel}},\ and\ \bibinfo {author} {\bibfnamefont
  {J.}~\bibnamefont {Wittmer}},\ }\bibfield  {title} {\bibinfo {title}
  {Strictly two-dimensional self-avoiding walks: Density crossover scaling},\
  }\href@noop {} {\bibfield  {journal} {\bibinfo  {journal} {Polymer Science
  Series C}\ }\textbf {\bibinfo {volume} {55}},\ \bibinfo {pages} {181}
  (\bibinfo {year} {2013})}\BibitemShut {NoStop}%
\bibitem [{\citenamefont {Liu}\ and\ \citenamefont
  {Hyeon}(2022)}]{Liu2022JPCB}%
  \BibitemOpen
  \bibfield  {author} {\bibinfo {author} {\bibfnamefont {L.}~\bibnamefont
  {Liu}}\ and\ \bibinfo {author} {\bibfnamefont {C.}~\bibnamefont {Hyeon}},\
  }\bibfield  {title} {\bibinfo {title} {Solvent quality dependent osmotic
  pressure of polymer solutions in two dimensions},\ }\href@noop {} {\bibfield
  {journal} {\bibinfo  {journal} {J. Phys. Chem. B}\ }\textbf {\bibinfo
  {volume} {126}},\ \bibinfo {pages} {9695} (\bibinfo {year}
  {2022})}\BibitemShut {NoStop}%
\bibitem [{\citenamefont {Mandelbrot}(1982)}]{mandelbrot1982fractal}%
  \BibitemOpen
  \bibfield  {author} {\bibinfo {author} {\bibfnamefont {B.~B.}\ \bibnamefont
  {Mandelbrot}},\ }\href@noop {} {\emph {\bibinfo {title} {The fractal geometry
  of nature}}},\ Vol.~\bibinfo {volume} {1}\ (\bibinfo  {publisher} {WH freeman
  New York},\ \bibinfo {year} {1982})\BibitemShut {NoStop}%
\bibitem [{\citenamefont {Coniglio}\ \emph {et~al.}(1987)\citenamefont
  {Coniglio}, \citenamefont {Jan}, \citenamefont {Majid},\ and\ \citenamefont
  {Stanley}}]{coniglio1987PRB}%
  \BibitemOpen
  \bibfield  {author} {\bibinfo {author} {\bibfnamefont {A.}~\bibnamefont
  {Coniglio}}, \bibinfo {author} {\bibfnamefont {N.}~\bibnamefont {Jan}},
  \bibinfo {author} {\bibfnamefont {I.}~\bibnamefont {Majid}},\ and\ \bibinfo
  {author} {\bibfnamefont {H.~E.}\ \bibnamefont {Stanley}},\ }\bibfield
  {title} {\bibinfo {title} {{Conformation of a polymer chain at the $\theta'$
  point: Connection to the external perimeter of a percolation cluster}},\
  }\href@noop {} {\bibfield  {journal} {\bibinfo  {journal} {Phys. Rev. B}\
  }\textbf {\bibinfo {volume} {35}},\ \bibinfo {pages} {3617} (\bibinfo {year}
  {1987})}\BibitemShut {NoStop}%
\bibitem [{\citenamefont {Duplantier}(1987{\natexlab{c}})}]{Duplantier87JCP}%
  \BibitemOpen
  \bibfield  {author} {\bibinfo {author} {\bibfnamefont {B.}~\bibnamefont
  {Duplantier}},\ }\bibfield  {title} {\bibinfo {title} {Geometry of polymer
  chains near the theta-point and dimensional regularization},\ }\href@noop {}
  {\bibfield  {journal} {\bibinfo  {journal} {J. Chem. Phys.}\ }\textbf
  {\bibinfo {volume} {86}},\ \bibinfo {pages} {4233} (\bibinfo {year}
  {1987}{\natexlab{c}})}\BibitemShut {NoStop}%
\bibitem [{\citenamefont {Jung}\ \emph {et~al.}(2020)\citenamefont {Jung},
  \citenamefont {Hyeon},\ and\ \citenamefont {Ha}}]{Jung2020Macromol}%
  \BibitemOpen
  \bibfield  {author} {\bibinfo {author} {\bibfnamefont {Y.}~\bibnamefont
  {Jung}}, \bibinfo {author} {\bibfnamefont {C.}~\bibnamefont {Hyeon}},\ and\
  \bibinfo {author} {\bibfnamefont {B.-Y.}\ \bibnamefont {Ha}},\ }\bibfield
  {title} {\bibinfo {title} {Near-$\theta$ polymers in a cylindrical space},\
  }\href@noop {} {\bibfield  {journal} {\bibinfo  {journal} {Macromolecules}\
  }\textbf {\bibinfo {volume} {53}},\ \bibinfo {pages} {2412} (\bibinfo {year}
  {2020})}\BibitemShut {NoStop}%
\bibitem [{\citenamefont {Beckrich}\ \emph {et~al.}(2007)\citenamefont
  {Beckrich}, \citenamefont {Johner}, \citenamefont {Semenov}, \citenamefont
  {Obukhov}, \citenamefont {Benoit},\ and\ \citenamefont
  {Wittmer}}]{beckrich2007macro}%
  \BibitemOpen
  \bibfield  {author} {\bibinfo {author} {\bibfnamefont {P.}~\bibnamefont
  {Beckrich}}, \bibinfo {author} {\bibfnamefont {A.}~\bibnamefont {Johner}},
  \bibinfo {author} {\bibfnamefont {A.~N.}\ \bibnamefont {Semenov}}, \bibinfo
  {author} {\bibfnamefont {S.~P.}\ \bibnamefont {Obukhov}}, \bibinfo {author}
  {\bibfnamefont {H.}~\bibnamefont {Benoit}},\ and\ \bibinfo {author}
  {\bibfnamefont {J.}~\bibnamefont {Wittmer}},\ }\bibfield  {title} {\bibinfo
  {title} {{Intramolecular form factor in dense polymer systems: Systematic
  deviations from the Debye formula}},\ }\href@noop {} {\bibfield  {journal}
  {\bibinfo  {journal} {Macromolecules}\ }\textbf {\bibinfo {volume} {40}},\
  \bibinfo {pages} {3805} (\bibinfo {year} {2007})}\BibitemShut {NoStop}%
\bibitem [{\citenamefont {Kager}\ and\ \citenamefont
  {Nienhuis}(2004)}]{kager2004guide}%
  \BibitemOpen
  \bibfield  {author} {\bibinfo {author} {\bibfnamefont {W.}~\bibnamefont
  {Kager}}\ and\ \bibinfo {author} {\bibfnamefont {B.}~\bibnamefont
  {Nienhuis}},\ }\bibfield  {title} {\bibinfo {title} {A guide to stochastic
  l{\"o}wner evolution and its applications},\ }\href@noop {} {\bibfield
  {journal} {\bibinfo  {journal} {J. Stat. Phys.}\ }\textbf {\bibinfo {volume}
  {115}},\ \bibinfo {pages} {1149} (\bibinfo {year} {2004})}\BibitemShut
  {NoStop}%
\bibitem [{\citenamefont {Rohde}\ and\ \citenamefont
  {Schramm}(2005)}]{rohde2005basic}%
  \BibitemOpen
  \bibfield  {author} {\bibinfo {author} {\bibfnamefont {S.}~\bibnamefont
  {Rohde}}\ and\ \bibinfo {author} {\bibfnamefont {O.}~\bibnamefont
  {Schramm}},\ }\bibfield  {title} {\bibinfo {title} {{Basic properties of
  SLE}},\ }\href@noop {} {\bibfield  {journal} {\bibinfo  {journal} {Ann.
  Math.}\ ,\ \bibinfo {pages} {883}} (\bibinfo {year} {2005})}\BibitemShut
  {NoStop}%
\bibitem [{\citenamefont {Cardy}(2005)}]{cardy2005sle}%
  \BibitemOpen
  \bibfield  {author} {\bibinfo {author} {\bibfnamefont {J.}~\bibnamefont
  {Cardy}},\ }\bibfield  {title} {\bibinfo {title} {{SLE for theoretical
  physicists}},\ }\href@noop {} {\bibfield  {journal} {\bibinfo  {journal}
  {Ann. Phys.}\ }\textbf {\bibinfo {volume} {318}},\ \bibinfo {pages} {81}
  (\bibinfo {year} {2005})}\BibitemShut {NoStop}%
\bibitem [{\citenamefont {Lawler}(2009)}]{lawler2009conformal}%
  \BibitemOpen
  \bibfield  {author} {\bibinfo {author} {\bibfnamefont {G.}~\bibnamefont
  {Lawler}},\ }\bibfield  {title} {\bibinfo {title} {Conformal invariance and
  2d statistical physics},\ }\href@noop {} {\bibfield  {journal} {\bibinfo
  {journal} {Bull. Am. Math. Soc.}\ }\textbf {\bibinfo {volume} {46}},\
  \bibinfo {pages} {35} (\bibinfo {year} {2009})}\BibitemShut {NoStop}%
\bibitem [{\citenamefont {Limbach}\ \emph {et~al.}(2006)\citenamefont
  {Limbach}, \citenamefont {Arnold}, \citenamefont {Mann},\ and\ \citenamefont
  {Holm}}]{limbach2006espresso}%
  \BibitemOpen
  \bibfield  {author} {\bibinfo {author} {\bibfnamefont {H.-J.}\ \bibnamefont
  {Limbach}}, \bibinfo {author} {\bibfnamefont {A.}~\bibnamefont {Arnold}},
  \bibinfo {author} {\bibfnamefont {B.~A.}\ \bibnamefont {Mann}},\ and\
  \bibinfo {author} {\bibfnamefont {C.}~\bibnamefont {Holm}},\ }\bibfield
  {title} {\bibinfo {title} {{ESPResSo—an extensible simulation package for
  research on soft matter systems}},\ }\href@noop {} {\bibfield  {journal}
  {\bibinfo  {journal} {Comp. Phys. Comm.}\ }\textbf {\bibinfo {volume}
  {174}},\ \bibinfo {pages} {704} (\bibinfo {year} {2006})}\BibitemShut
  {NoStop}%
\bibitem [{\citenamefont {Saberi}(2009)}]{saberi2009thermal}%
  \BibitemOpen
  \bibfield  {author} {\bibinfo {author} {\bibfnamefont {A.~A.}\ \bibnamefont
  {Saberi}},\ }\bibfield  {title} {\bibinfo {title} {{Thermal behavior of spin
  clusters and interfaces in the two-dimensional Ising model on a square
  lattice}},\ }\href@noop {} {\bibfield  {journal} {\bibinfo  {journal} {J.
  Stat. Mech.: Theor. Exp.}\ }\textbf {\bibinfo {volume} {2009}},\ \bibinfo
  {pages} {P07030} (\bibinfo {year} {2009})}\BibitemShut {NoStop}%
\bibitem [{\citenamefont {Saberi}(2011)}]{Saberi2011PRE}%
  \BibitemOpen
  \bibfield  {author} {\bibinfo {author} {\bibfnamefont {A.~A.}\ \bibnamefont
  {Saberi}},\ }\bibfield  {title} {\bibinfo {title} {Fractal structure of a
  three-dimensional brownian motion on an attractive plane},\ }\href@noop {}
  {\bibfield  {journal} {\bibinfo  {journal} {Phys. Rev. E}\ }\textbf {\bibinfo
  {volume} {84}},\ \bibinfo {pages} {021113} (\bibinfo {year}
  {2011})}\BibitemShut {NoStop}%
\bibitem [{\citenamefont {Lawler}\ \emph {et~al.}(2000)\citenamefont {Lawler},
  \citenamefont {Schramm},\ and\ \citenamefont {Werner}}]{Lawler2000TheDO}%
  \BibitemOpen
  \bibfield  {author} {\bibinfo {author} {\bibfnamefont {G.~F.}\ \bibnamefont
  {Lawler}}, \bibinfo {author} {\bibfnamefont {O.}~\bibnamefont {Schramm}},\
  and\ \bibinfo {author} {\bibfnamefont {W.}~\bibnamefont {Werner}},\
  }\bibfield  {title} {\bibinfo {title} {The dimension of the planar brownian
  frontier is 4/3},\ }\href@noop {} {\bibfield  {journal} {\bibinfo  {journal}
  {Math. Res. Lett.}\ }\textbf {\bibinfo {volume} {8}},\ \bibinfo {pages} {401}
  (\bibinfo {year} {2000})}\BibitemShut {NoStop}%
\bibitem [{\citenamefont {Sapoval}\ \emph {et~al.}(1985)\citenamefont
  {Sapoval}, \citenamefont {Rosso},\ and\ \citenamefont
  {Gouyet}}]{Sapoval85JPL}%
  \BibitemOpen
  \bibfield  {author} {\bibinfo {author} {\bibfnamefont {B.}~\bibnamefont
  {Sapoval}}, \bibinfo {author} {\bibfnamefont {M.}~\bibnamefont {Rosso}},\
  and\ \bibinfo {author} {\bibfnamefont {F.}~\bibnamefont {Gouyet}},\
  }\bibfield  {title} {\bibinfo {title} {The fractal nature of a diffusion
  front and the relation to percolation},\ }\href@noop {} {\bibfield  {journal}
  {\bibinfo  {journal} {J. Phys. Lett.}\ }\textbf {\bibinfo {volume} {46}},\
  \bibinfo {pages} {L149} (\bibinfo {year} {1985})}\BibitemShut {NoStop}%
\bibitem [{\citenamefont {Bunde}\ and\ \citenamefont
  {Gouyet}(1985)}]{Bunde85JPA}%
  \BibitemOpen
  \bibfield  {author} {\bibinfo {author} {\bibfnamefont {A.}~\bibnamefont
  {Bunde}}\ and\ \bibinfo {author} {\bibfnamefont {J.~F.}\ \bibnamefont
  {Gouyet}},\ }\bibfield  {title} {\bibinfo {title} {On scaling relations in
  growth models for percolating clusters and diffusion fronts},\ }\href@noop {}
  {\bibfield  {journal} {\bibinfo  {journal} {J. Phys. A: Math. Gen.}\ }\textbf
  {\bibinfo {volume} {18}},\ \bibinfo {pages} {L2850} (\bibinfo {year}
  {1985})}\BibitemShut {NoStop}%
\bibitem [{\citenamefont {Duplantier}\ and\ \citenamefont
  {Saleur}(1989)}]{duplantier1989PRL}%
  \BibitemOpen
  \bibfield  {author} {\bibinfo {author} {\bibfnamefont {B.}~\bibnamefont
  {Duplantier}}\ and\ \bibinfo {author} {\bibfnamefont {H.}~\bibnamefont
  {Saleur}},\ }\bibfield  {title} {\bibinfo {title} {{Stability of the polymer
  $\Theta$ point in two dimensions}},\ }\href@noop {} {\bibfield  {journal}
  {\bibinfo  {journal} {Phys. Rev. Lett.}\ }\textbf {\bibinfo {volume} {62}},\
  \bibinfo {pages} {1368} (\bibinfo {year} {1989})}\BibitemShut {NoStop}%
\bibitem [{\citenamefont {Grossman}\ and\ \citenamefont
  {Aharony}(1986)}]{grossman1986structure}%
  \BibitemOpen
  \bibfield  {author} {\bibinfo {author} {\bibfnamefont {T.}~\bibnamefont
  {Grossman}}\ and\ \bibinfo {author} {\bibfnamefont {A.}~\bibnamefont
  {Aharony}},\ }\bibfield  {title} {\bibinfo {title} {Structure and perimeters
  of percolation clusters},\ }\href@noop {} {\bibfield  {journal} {\bibinfo
  {journal} {J. Phys. A: Math.}\ }\textbf {\bibinfo {volume} {19}},\ \bibinfo
  {pages} {L745} (\bibinfo {year} {1986})}\BibitemShut {NoStop}%
\bibitem [{\citenamefont {Grossman}\ and\ \citenamefont
  {Aharony}(1987)}]{grossman1987accessible}%
  \BibitemOpen
  \bibfield  {author} {\bibinfo {author} {\bibfnamefont {T.}~\bibnamefont
  {Grossman}}\ and\ \bibinfo {author} {\bibfnamefont {A.}~\bibnamefont
  {Aharony}},\ }\bibfield  {title} {\bibinfo {title} {Accessible external
  perimeters of percolation clusters},\ }\href@noop {} {\bibfield  {journal}
  {\bibinfo  {journal} {J. Phys. A: Math. Gen.}\ }\textbf {\bibinfo {volume}
  {20}},\ \bibinfo {pages} {L1193} (\bibinfo {year} {1987})}\BibitemShut
  {NoStop}%
\bibitem [{\citenamefont {Aizenman}\ \emph {et~al.}(1999)\citenamefont
  {Aizenman}, \citenamefont {Duplantier},\ and\ \citenamefont
  {Aharony}}]{aizenman1999PRL}%
  \BibitemOpen
  \bibfield  {author} {\bibinfo {author} {\bibfnamefont {M.}~\bibnamefont
  {Aizenman}}, \bibinfo {author} {\bibfnamefont {B.}~\bibnamefont
  {Duplantier}},\ and\ \bibinfo {author} {\bibfnamefont {A.}~\bibnamefont
  {Aharony}},\ }\bibfield  {title} {\bibinfo {title} {{Path-crossing exponents
  and the external perimeter in 2D percolation}},\ }\href@noop {} {\bibfield
  {journal} {\bibinfo  {journal} {Phys. Rev. Lett.}\ }\textbf {\bibinfo
  {volume} {83}},\ \bibinfo {pages} {1359} (\bibinfo {year}
  {1999})}\BibitemShut {NoStop}%
\bibitem [{\citenamefont {Duplantier}(1986{\natexlab{b}})}]{duplantier1986PRL}%
  \BibitemOpen
  \bibfield  {author} {\bibinfo {author} {\bibfnamefont {B.}~\bibnamefont
  {Duplantier}},\ }\bibfield  {title} {\bibinfo {title} {{Polymer network of
  fixed topology: renormalization, exact critical exponent $\gamma$ in two
  dimensions, and $d= 4- \varepsilon$}},\ }\href@noop {} {\bibfield  {journal}
  {\bibinfo  {journal} {Phys. Rev. Lett.}\ }\textbf {\bibinfo {volume} {57}},\
  \bibinfo {pages} {941} (\bibinfo {year} {1986}{\natexlab{b}})}\BibitemShut
  {NoStop}%
\bibitem [{\citenamefont {Kennedy}(2009)}]{kennedy2009numerical}%
  \BibitemOpen
  \bibfield  {author} {\bibinfo {author} {\bibfnamefont {T.}~\bibnamefont
  {Kennedy}},\ }\bibfield  {title} {\bibinfo {title} {{Numerical computations
  for the Schramm-Loewner evolution}},\ }\href@noop {} {\bibfield  {journal}
  {\bibinfo  {journal} {J. Stat. Phys.}\ }\textbf {\bibinfo {volume} {137}},\
  \bibinfo {pages} {839} (\bibinfo {year} {2009})}\BibitemShut {NoStop}%
\bibitem [{\citenamefont {L{\"o}wner}(1923)}]{lowner1923untersuchungen}%
  \BibitemOpen
  \bibfield  {author} {\bibinfo {author} {\bibfnamefont {K.}~\bibnamefont
  {L{\"o}wner}},\ }\bibfield  {title} {\bibinfo {title} {{Untersuchungen
  {\"u}ber schlichte konforme Abbildungen des Einheitskreises. I}},\
  }\href@noop {} {\bibfield  {journal} {\bibinfo  {journal} {Mathematische
  Annalen}\ }\textbf {\bibinfo {volume} {89}},\ \bibinfo {pages} {103}
  (\bibinfo {year} {1923})}\BibitemShut {NoStop}%
\bibitem [{\citenamefont {Beffara}(2008)}]{beffara2008dimension}%
  \BibitemOpen
  \bibfield  {author} {\bibinfo {author} {\bibfnamefont {V.}~\bibnamefont
  {Beffara}},\ }\bibfield  {title} {\bibinfo {title} {{The dimension of the SLE
  curves}},\ }\href@noop {} {\bibfield  {journal} {\bibinfo  {journal} {Ann.
  Probab.}\ }\textbf {\bibinfo {volume} {36}},\ \bibinfo {pages} {1421}
  (\bibinfo {year} {2008})}\BibitemShut {NoStop}%
\bibitem [{\citenamefont {Duplantier}(2000)}]{duplantier2000conformally}%
  \BibitemOpen
  \bibfield  {author} {\bibinfo {author} {\bibfnamefont {B.}~\bibnamefont
  {Duplantier}},\ }\bibfield  {title} {\bibinfo {title} {Conformally invariant
  fractals and potential theory},\ }\href@noop {} {\bibfield  {journal}
  {\bibinfo  {journal} {Phys. Rev. Lett.}\ }\textbf {\bibinfo {volume} {84}},\
  \bibinfo {pages} {1363} (\bibinfo {year} {2000})}\BibitemShut {NoStop}%
\bibitem [{\citenamefont {Bauer}\ and\ \citenamefont
  {Bernard}(2006)}]{bauer20062d}%
  \BibitemOpen
  \bibfield  {author} {\bibinfo {author} {\bibfnamefont {M.}~\bibnamefont
  {Bauer}}\ and\ \bibinfo {author} {\bibfnamefont {D.}~\bibnamefont
  {Bernard}},\ }\bibfield  {title} {\bibinfo {title} {{2D growth processes: SLE
  and Loewner chains}},\ }\href@noop {} {\bibfield  {journal} {\bibinfo
  {journal} {Phys. Rep.}\ }\textbf {\bibinfo {volume} {432}},\ \bibinfo {pages}
  {115} (\bibinfo {year} {2006})}\BibitemShut {NoStop}%
\bibitem [{\citenamefont {Qin}\ \emph {et~al.}(2021)\citenamefont {Qin},
  \citenamefont {Fei}, \citenamefont {Wang}, \citenamefont {Stone},
  \citenamefont {Wingreen},\ and\ \citenamefont
  {Bassler}}]{qin2021hierarchical}%
  \BibitemOpen
  \bibfield  {author} {\bibinfo {author} {\bibfnamefont {B.}~\bibnamefont
  {Qin}}, \bibinfo {author} {\bibfnamefont {C.}~\bibnamefont {Fei}}, \bibinfo
  {author} {\bibfnamefont {B.}~\bibnamefont {Wang}}, \bibinfo {author}
  {\bibfnamefont {H.~A.}\ \bibnamefont {Stone}}, \bibinfo {author}
  {\bibfnamefont {N.~S.}\ \bibnamefont {Wingreen}},\ and\ \bibinfo {author}
  {\bibfnamefont {B.~L.}\ \bibnamefont {Bassler}},\ }\bibfield  {title}
  {\bibinfo {title} {Hierarchical transitions and fractal wrinkling drive
  bacterial pellicle morphogenesis},\ }\href@noop {} {\bibfield  {journal}
  {\bibinfo  {journal} {Proc. Natl. Acad. Sci. USA}\ }\textbf {\bibinfo
  {volume} {118}},\ \bibinfo {pages} {e2023504118} (\bibinfo {year}
  {2021})}\BibitemShut {NoStop}%
\bibitem [{\citenamefont {Brooks}\ \emph {et~al.}(2022)\citenamefont {Brooks},
  \citenamefont {McCool}, \citenamefont {Gillman}, \citenamefont {Suel},
  \citenamefont {Mugler},\ and\ \citenamefont
  {Larkin}}]{brooks2022computational}%
  \BibitemOpen
  \bibfield  {author} {\bibinfo {author} {\bibfnamefont {C.}~\bibnamefont
  {Brooks}}, \bibinfo {author} {\bibfnamefont {J.~T.}\ \bibnamefont {McCool}},
  \bibinfo {author} {\bibfnamefont {A.}~\bibnamefont {Gillman}}, \bibinfo
  {author} {\bibfnamefont {G.~M.}\ \bibnamefont {Suel}}, \bibinfo {author}
  {\bibfnamefont {A.}~\bibnamefont {Mugler}},\ and\ \bibinfo {author}
  {\bibfnamefont {J.~W.}\ \bibnamefont {Larkin}},\ }\bibfield  {title}
  {\bibinfo {title} {Computational model of fractal interface formation in
  bacterial biofilms},\ }\bibfield  {journal} {\bibinfo  {journal} {bioRxiv}\
  }\href {https://doi.org/https://doi.org/10.1101/2022.05.10.491419}
  {https://doi.org/10.1101/2022.05.10.491419} (\bibinfo {year}
  {2022})\BibitemShut {NoStop}%
\bibitem [{\citenamefont {Vilanove}\ and\ \citenamefont
  {Rondelez}(1980)}]{vilanove1980PRL}%
  \BibitemOpen
  \bibfield  {author} {\bibinfo {author} {\bibfnamefont {R.}~\bibnamefont
  {Vilanove}}\ and\ \bibinfo {author} {\bibfnamefont {F.}~\bibnamefont
  {Rondelez}},\ }\bibfield  {title} {\bibinfo {title} {Scaling description of
  two-dimensional chain conformations in polymer monolayers},\ }\href@noop {}
  {\bibfield  {journal} {\bibinfo  {journal} {Phys. Rev. Lett.}\ }\textbf
  {\bibinfo {volume} {45}},\ \bibinfo {pages} {1502} (\bibinfo {year}
  {1980})}\BibitemShut {NoStop}%
\bibitem [{\citenamefont {Witte}\ \emph {et~al.}(2010)\citenamefont {Witte},
  \citenamefont {Kewalramani}, \citenamefont {Kuzmenko}, \citenamefont {Sun},
  \citenamefont {Fukuto},\ and\ \citenamefont {Won}}]{witte2010macromolecules}%
  \BibitemOpen
  \bibfield  {author} {\bibinfo {author} {\bibfnamefont {K.~N.}\ \bibnamefont
  {Witte}}, \bibinfo {author} {\bibfnamefont {S.}~\bibnamefont {Kewalramani}},
  \bibinfo {author} {\bibfnamefont {I.}~\bibnamefont {Kuzmenko}}, \bibinfo
  {author} {\bibfnamefont {W.}~\bibnamefont {Sun}}, \bibinfo {author}
  {\bibfnamefont {M.}~\bibnamefont {Fukuto}},\ and\ \bibinfo {author}
  {\bibfnamefont {Y.-Y.}\ \bibnamefont {Won}},\ }\bibfield  {title} {\bibinfo
  {title} {Formation and collapse of single-monomer-thick monolayers of poly
  (n-butyl acrylate) at the air-water interface},\ }\href@noop {} {\bibfield
  {journal} {\bibinfo  {journal} {Macromolecules}\ }\textbf {\bibinfo {volume}
  {43}},\ \bibinfo {pages} {2990} (\bibinfo {year} {2010})}\BibitemShut
  {NoStop}%
\bibitem [{\citenamefont {Goicochea}\ and\ \citenamefont
  {P{\'e}rez}(2015)}]{goicochea2015scaling}%
  \BibitemOpen
  \bibfield  {author} {\bibinfo {author} {\bibfnamefont {A.~G.}\ \bibnamefont
  {Goicochea}}\ and\ \bibinfo {author} {\bibfnamefont {E.}~\bibnamefont
  {P{\'e}rez}},\ }\bibfield  {title} {\bibinfo {title} {Scaling law of the
  disjoining pressure reveals 2d structure of polymeric fluids},\ }\href@noop
  {} {\bibfield  {journal} {\bibinfo  {journal} {Macromol. Chem. Phys.}\
  }\textbf {\bibinfo {volume} {216}},\ \bibinfo {pages} {1076} (\bibinfo {year}
  {2015})}\BibitemShut {NoStop}%
\bibitem [{\citenamefont {Jones}\ \emph {et~al.}(1999)\citenamefont {Jones},
  \citenamefont {Kumar}, \citenamefont {Ho}, \citenamefont {Briber},\ and\
  \citenamefont {Russell}}]{jones1999chain}%
  \BibitemOpen
  \bibfield  {author} {\bibinfo {author} {\bibfnamefont {R.~L.}\ \bibnamefont
  {Jones}}, \bibinfo {author} {\bibfnamefont {S.~K.}\ \bibnamefont {Kumar}},
  \bibinfo {author} {\bibfnamefont {D.~L.}\ \bibnamefont {Ho}}, \bibinfo
  {author} {\bibfnamefont {R.~M.}\ \bibnamefont {Briber}},\ and\ \bibinfo
  {author} {\bibfnamefont {T.~P.}\ \bibnamefont {Russell}},\ }\bibfield
  {title} {\bibinfo {title} {Chain conformation in ultrathin polymer films},\
  }\href@noop {} {\bibfield  {journal} {\bibinfo  {journal} {Nature}\ }\textbf
  {\bibinfo {volume} {400}},\ \bibinfo {pages} {146} (\bibinfo {year}
  {1999})}\BibitemShut {NoStop}%
\bibitem [{\citenamefont {Kumaki}\ \emph {et~al.}(2010)\citenamefont {Kumaki},
  \citenamefont {Kajitani}, \citenamefont {Nagai}, \citenamefont {Okoshi},\
  and\ \citenamefont {Yashima}}]{kumaki2010visualization}%
  \BibitemOpen
  \bibfield  {author} {\bibinfo {author} {\bibfnamefont {J.}~\bibnamefont
  {Kumaki}}, \bibinfo {author} {\bibfnamefont {T.}~\bibnamefont {Kajitani}},
  \bibinfo {author} {\bibfnamefont {K.}~\bibnamefont {Nagai}}, \bibinfo
  {author} {\bibfnamefont {K.}~\bibnamefont {Okoshi}},\ and\ \bibinfo {author}
  {\bibfnamefont {E.}~\bibnamefont {Yashima}},\ }\bibfield  {title} {\bibinfo
  {title} {Visualization of polymer chain conformations in amorphous
  polyisocyanide langmuir- blodgett films by atomic force microscopy},\
  }\href@noop {} {\bibfield  {journal} {\bibinfo  {journal} {J. Am. Chem.
  Soc.}\ }\textbf {\bibinfo {volume} {132}},\ \bibinfo {pages} {5604} (\bibinfo
  {year} {2010})}\BibitemShut {NoStop}%
\bibitem [{\citenamefont {Sugihara}\ and\ \citenamefont
  {Kumaki}(2012)}]{sugihara2012visualization}%
  \BibitemOpen
  \bibfield  {author} {\bibinfo {author} {\bibfnamefont {K.}~\bibnamefont
  {Sugihara}}\ and\ \bibinfo {author} {\bibfnamefont {J.}~\bibnamefont
  {Kumaki}},\ }\bibfield  {title} {\bibinfo {title} {Visualization of
  two-dimensional single chain conformations solubilized in a miscible polymer
  blend monolayer by atomic force microscopy},\ }\href@noop {} {\bibfield
  {journal} {\bibinfo  {journal} {J. Phys. Chem. B}\ }\textbf {\bibinfo
  {volume} {116}},\ \bibinfo {pages} {6561} (\bibinfo {year}
  {2012})}\BibitemShut {NoStop}%
\bibitem [{\citenamefont {O'connell}\ and\ \citenamefont
  {McKenna}(2005)}]{o2005rheological}%
  \BibitemOpen
  \bibfield  {author} {\bibinfo {author} {\bibfnamefont {P.}~\bibnamefont
  {O'connell}}\ and\ \bibinfo {author} {\bibfnamefont {G.}~\bibnamefont
  {McKenna}},\ }\bibfield  {title} {\bibinfo {title} {Rheological measurements
  of the thermoviscoelastic response of ultrathin polymer films},\ }\href@noop
  {} {\bibfield  {journal} {\bibinfo  {journal} {Science}\ }\textbf {\bibinfo
  {volume} {307}},\ \bibinfo {pages} {1760} (\bibinfo {year}
  {2005})}\BibitemShut {NoStop}%
\bibitem [{\citenamefont {Wittmer}\ \emph {et~al.}(2010)\citenamefont
  {Wittmer}, \citenamefont {Meyer}, \citenamefont {Johner}, \citenamefont
  {Kreer},\ and\ \citenamefont {Baschnagel}}]{wittmer2010algebraic}%
  \BibitemOpen
  \bibfield  {author} {\bibinfo {author} {\bibfnamefont {J.}~\bibnamefont
  {Wittmer}}, \bibinfo {author} {\bibfnamefont {H.}~\bibnamefont {Meyer}},
  \bibinfo {author} {\bibfnamefont {A.}~\bibnamefont {Johner}}, \bibinfo
  {author} {\bibfnamefont {T.}~\bibnamefont {Kreer}},\ and\ \bibinfo {author}
  {\bibfnamefont {J.}~\bibnamefont {Baschnagel}},\ }\bibfield  {title}
  {\bibinfo {title} {Algebraic displacement correlation in two-dimensional
  polymer melts},\ }\href@noop {} {\bibfield  {journal} {\bibinfo  {journal}
  {Phys. Rev. Lett.}\ }\textbf {\bibinfo {volume} {105}},\ \bibinfo {pages}
  {037802} (\bibinfo {year} {2010})}\BibitemShut {NoStop}%
\bibitem [{\citenamefont {Nienhuis}(1982)}]{nienhuis1982exact}%
  \BibitemOpen
  \bibfield  {author} {\bibinfo {author} {\bibfnamefont {B.}~\bibnamefont
  {Nienhuis}},\ }\bibfield  {title} {\bibinfo {title} {{Exact critical point
  and critical exponents of $O(n)$ models in two dimensions}},\ }\href@noop {}
  {\bibfield  {journal} {\bibinfo  {journal} {Phys. Rev. Lett.}\ }\textbf
  {\bibinfo {volume} {49}},\ \bibinfo {pages} {1062} (\bibinfo {year}
  {1982})}\BibitemShut {NoStop}%
\bibitem [{\citenamefont {Duminil-Copin}\ and\ \citenamefont
  {Smirnov}(2012)}]{duminil2012connective}%
  \BibitemOpen
  \bibfield  {author} {\bibinfo {author} {\bibfnamefont {H.}~\bibnamefont
  {Duminil-Copin}}\ and\ \bibinfo {author} {\bibfnamefont {S.}~\bibnamefont
  {Smirnov}},\ }\bibfield  {title} {\bibinfo {title} {The connective constant
  of the honeycomb lattice equals $\sqrt{2+\sqrt{2}}$},\ }\href@noop {}
  {\bibfield  {journal} {\bibinfo  {journal} {Ann. Math.}\ }\textbf {\bibinfo
  {volume} {175}},\ \bibinfo {pages} {1653} (\bibinfo {year}
  {2012})}\BibitemShut {NoStop}%
\bibitem [{\citenamefont {{de Gennes}}(1972)}]{deGennes72PhysLett}%
  \BibitemOpen
  \bibfield  {author} {\bibinfo {author} {\bibfnamefont {P.~G.}\ \bibnamefont
  {{de Gennes}}},\ }\bibfield  {title} {\bibinfo {title} {{Exponents for the
  excluded volume problem as derived by the Wilson method}},\ }\href@noop {}
  {\bibfield  {journal} {\bibinfo  {journal} {Phys. Lett.}\ }\textbf {\bibinfo
  {volume} {38A}},\ \bibinfo {pages} {339} (\bibinfo {year}
  {1972})}\BibitemShut {NoStop}%
\bibitem [{\citenamefont {Nienhuis}(1984)}]{nienhuis1984critical}%
  \BibitemOpen
  \bibfield  {author} {\bibinfo {author} {\bibfnamefont {B.}~\bibnamefont
  {Nienhuis}},\ }\bibfield  {title} {\bibinfo {title} {{Critical behavior of
  two-dimensional spin models and charge asymmetry in the Coulomb gas}},\
  }\href@noop {} {\bibfield  {journal} {\bibinfo  {journal} {J. Stat. Phys.}\
  }\textbf {\bibinfo {volume} {34}},\ \bibinfo {pages} {731} (\bibinfo {year}
  {1984})}\BibitemShut {NoStop}%
\bibitem [{\citenamefont {Kosterlitz}\ and\ \citenamefont
  {Thouless}(1973)}]{kosterlitz1973ordering}%
  \BibitemOpen
  \bibfield  {author} {\bibinfo {author} {\bibfnamefont {J.~M.}\ \bibnamefont
  {Kosterlitz}}\ and\ \bibinfo {author} {\bibfnamefont {D.~J.}\ \bibnamefont
  {Thouless}},\ }\bibfield  {title} {\bibinfo {title} {Ordering, metastability
  and phase transitions in two-dimensional systems},\ }\href@noop {} {\bibfield
   {journal} {\bibinfo  {journal} {J. Phys. C: Solid State Physics}\ }\textbf
  {\bibinfo {volume} {6}},\ \bibinfo {pages} {1181} (\bibinfo {year}
  {1973})}\BibitemShut {NoStop}%
\bibitem [{\citenamefont {Kosterlitz}(1974)}]{kosterlitz1974critical}%
  \BibitemOpen
  \bibfield  {author} {\bibinfo {author} {\bibfnamefont {J.}~\bibnamefont
  {Kosterlitz}},\ }\bibfield  {title} {\bibinfo {title} {The critical
  properties of the two-dimensional xy model},\ }\href@noop {} {\bibfield
  {journal} {\bibinfo  {journal} {J. Phys. C Solid State Physics}\ }\textbf
  {\bibinfo {volume} {7}},\ \bibinfo {pages} {1046} (\bibinfo {year}
  {1974})}\BibitemShut {NoStop}%
\bibitem [{\citenamefont {Jos{\'e}}\ \emph {et~al.}(1977)\citenamefont
  {Jos{\'e}}, \citenamefont {Kadanoff}, \citenamefont {Kirkpatrick},\ and\
  \citenamefont {Nelson}}]{jose1977renormalization}%
  \BibitemOpen
  \bibfield  {author} {\bibinfo {author} {\bibfnamefont {J.~V.}\ \bibnamefont
  {Jos{\'e}}}, \bibinfo {author} {\bibfnamefont {L.~P.}\ \bibnamefont
  {Kadanoff}}, \bibinfo {author} {\bibfnamefont {S.}~\bibnamefont
  {Kirkpatrick}},\ and\ \bibinfo {author} {\bibfnamefont {D.~R.}\ \bibnamefont
  {Nelson}},\ }\bibfield  {title} {\bibinfo {title} {Renormalization, vortices,
  and symmetry-breaking perturbations in the two-dimensional planar model},\
  }\href@noop {} {\bibfield  {journal} {\bibinfo  {journal} {Phys. Rev. B}\
  }\textbf {\bibinfo {volume} {16}},\ \bibinfo {pages} {1217} (\bibinfo {year}
  {1977})}\BibitemShut {NoStop}%
\end{thebibliography}%

\end{document}